\documentclass[journal]{IEEEtran}

\usepackage[cmex10]{amsmath}
\usepackage{amssymb}

\usepackage[pdftex]{graphicx}
\graphicspath{{pics/}}

\usepackage{booktabs}

\begin{document}
%
\title{Measurement-Based Wideband Analysis of \\Dynamic  Multipath Propagation in \\Vehicular Communication Scenarios}
%
%
%

\author{Kim~Mahler,~Wilhelm~Keusgen,~Fredrik~Tufvesson,~Thomas~Zemen~and~Giuseppe~Caire
\thanks{Copyright \copyright~2016 IEEE. Personal use of this material is permitted. However, permission to use this material for any other purposes must be obtained from the IEEE by sending a request to pubs-permissions@ieee.org. }
\thanks{K. Mahler and W. Keusgen are with the Department of Wireless Communications and Networks, Fraunhofer Heinrich Hertz Institute, Berlin, Germany. e-mail: (kim.mahler@hhi.fraunhofer.de)}
\thanks{F. Tufvesson is with the Department of Electrical and Information Technology, Lund University, Lund, Sweden.}
\thanks{T. Zemen is with the Digital Safety and Security Department, AIT Austrian Institute of Technology, Vienna, Austria.}
\thanks{G. Caire is with the Communications and Information Theory Group, Technische Universit\"at Berlin, Berlin, Germany.}
\thanks{Manuscript received April 12th, 2016.}}

\maketitle

\begin{abstract}
Realistic propagation modeling requires a detailed understanding and characterization of the radio channel properties. This paper is based on channel sounder measurements with 1 GHz bandwidth at a carrier frequency of 5.7 GHz and particular tracking methods. We present statistical models for the number of, birth rate, lifetime, excess delay and relative Doppler frequency of individual multipath components (MPCs). Our findings are concluded from 72 measurement runs in eight relevant vehicular communication scenarios and reveal wide insights into the dynamic propagation process in vehicular communication scenarios.
\end{abstract}


\IEEEpeerreviewmaketitle
\section{Introduction}

\IEEEPARstart {V}{ehicular} communication will play an important role in future safety-related traffic applications, since it enables direct information exchange between vehicles without line-of-sight. Highly reliable wireless communication links are key components for these applications. Those are in turn directly dependent on the propagation channel, the received power and the destructive fading effects of the channel. To get a better understanding of the multipath process in vehicular communication scenarios, thorough analysis of measurement data from channel sounding campaigns is necessary. Based on the findings, suitable channel models with appropriate parametrization can be developed and used for advanced transceiver development towards the desired reliability levels.  

Communication systems based on IEEE~802.11p are close to market launch and operate at 5.9~GHz. Therefore, most researchers in the field conducted vehicular channel measurements in the 6 GHz band with measurement bandwidths of 60-240~MHz \cite{renaudin10} \cite{karedal10}. The main advantages of a wider measurement bandwidth are the high delay time resolution of individual multipath components (MPC) and the reduced sensitivity to small-scale fading effects, due to a lower number of superimposed MPCs per delay time bin \cite{molisch05}. Another benefit of highly resolved wideband channel data is the ability to relate individual MPCs to physical scattering objects.
In terms of vehicle-to-vehicle (V2V) channel modeling, there are two major paths being followed: 1) the tapped delay line (TDL) model and its extensions or derivatives and 2) geometry-based modeling approaches. An interesting comparison of these approaches is given in \cite{Mat11}, different geometry-based approaches are presented in \cite{Wang09}. Different from these empirical modeling approaches, V2V channel models are analytically derived from the geometry in \cite{Patz1} \cite{Patz2} \cite{Zaj1} \cite{Zaj2}. Also, analytical regular-shaped geometry-based stochastic models are introduced in \cite{Cheng1} \cite{Cheng2}. Non-stationary channels are also studied in \cite{Patz3} \cite{Patz4} \cite{Patz5}. 

Probably the two most distinctive characteristics of V2V channels are the low position of the antennas and the time-variant radio properties, due to the movement of transmitter, receiver and scattering objects. These characteristics lead to, among others, fast MPC shadowing effects and diverse Doppler frequency distributions \cite{bernado14}. Hence, classical TDL models are enhanced with a ''persistence process'' \cite{Sen08} or with varying Doppler spectra models \cite{Aco07}. Different from these narrowband models, \cite{he15} proposes a dynamic wideband V2V channel model based on a local wide-sense stationary (WSS) time window and MPC statistics related to this time window. 
Alternatively, V2V geometry-based stochastic channel models were proposed, which are well-suited for non-stationary environments \cite{karedal09}. In order to identify the time-variant stochastics of these V2V channel models, researchers focus on the temporal behavior of MPCs. Work in \cite{Ming13} \cite{Xu15} investigate MPC clusters in the delay-Doppler plane and model their temporal behavior. To the author's knowledge, no publication provides a substantial characterization of individual MPCs and their dynamics in various vehicular communication scenarios.

The contribution of this paper is a comprehensive analysis of an extensive channel data set and a statistical characterization of the dynamic behavior of individual wideband (1~GHz) MPC in V2V scenarios, i.e. the number, birthrate and lifetime of MPCs. Statistical analysis of individual MPC delay and Doppler observations are also given and appropriate models for all key channel parameters proposed. The analysis is based on 72 measurement runs in eight different vehicular communication scenarios. The results of this paper provide wide insights into the vehicular propagation channel and are beneficial for the parametrization of various V2V channel modeling approaches.

The paper is organized as follows: we describe the measurement equipment, settings and environments of the vehicular communication scenarios in Section II. The applied extraction methods for the analyzed channel properties are explained in Section III. The results of our analysis are presented in Section IV, subdivided into the five investigated channel parameters. The paper closes with conclusions in Section V.

\section{Measurement}
\subsection{Measurement Equipment}
\begin{figure}[!t]
\centering
\includegraphics[width=3.3in]{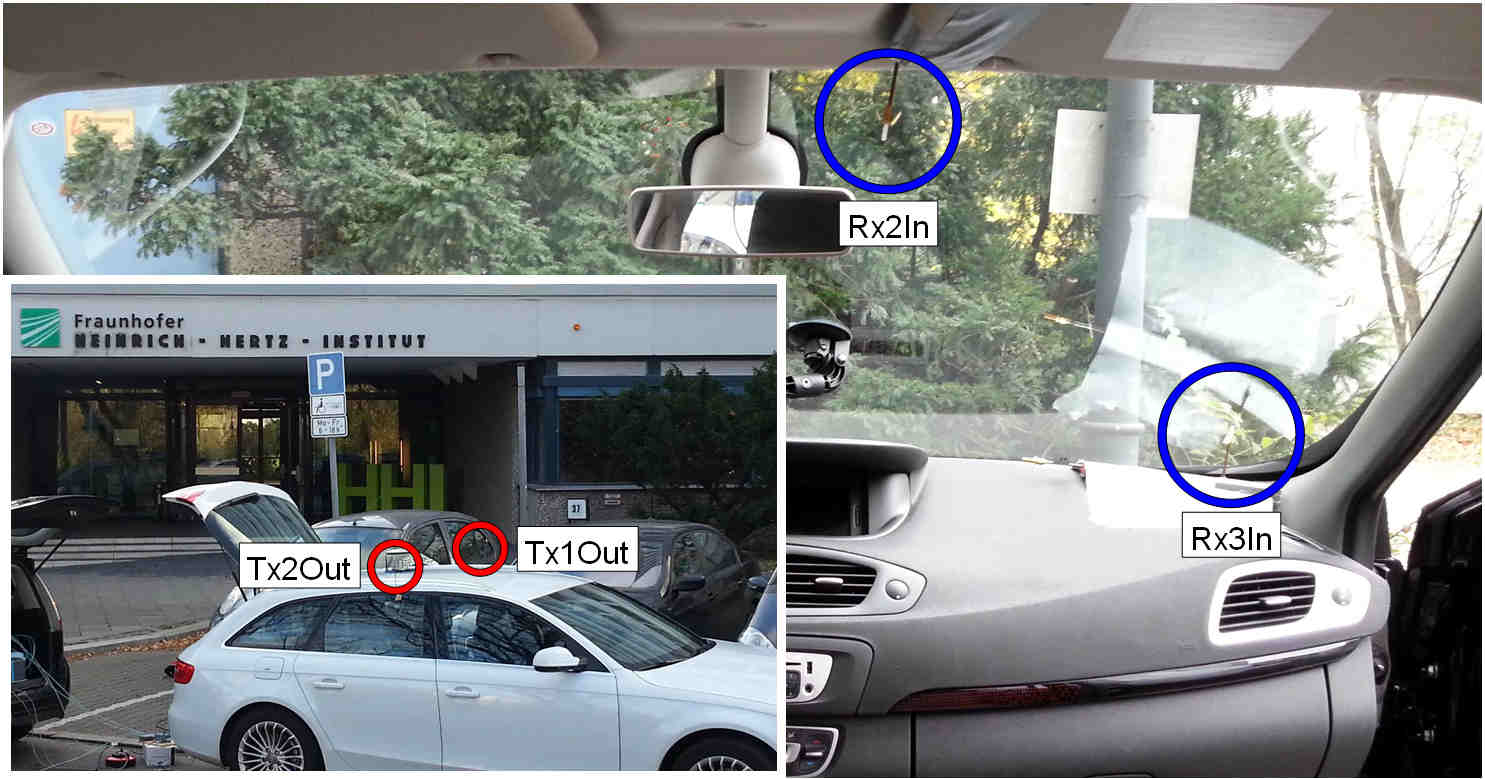}
\caption{Position of antennas during measurement: two Tx and three Rx antennas (Rx1Out is in a similar position as Tx2Out).}
\label{antenna}
\end{figure}

The HHI channel sounder, developed at the Fraunhofer Heinrich Hertz Institute (HHI), is a wideband measurement device with a bandwidth of 1~GHz at a carrier frequency of 5.7~GHz \cite{paschalidis08}. The measurement bandwidth permits a delay time resolution of 1~ns (30~cm of wave propagation) and therefore a highly resolved view into the behavior of MPCs. The channel sounder consists of a transmitter unit and a receiver unit that can be installed in conventional passenger vehicles and deployed in real traffic scenarios. For our measurements we use an Audi A4 Avant as transmitter and a Renault Scenic as receiver vehicle. The transmitter vehicle is equipped with two omnidirectional and vertically polarized antennas mounted on the roof at the left and right edges of the vehicle (Tx1Out and Tx2Out in Fig. \ref{antenna}). The receiver vehicle is also equipped with the same kind of antenna on the roof at the left edge of the vehicle (Rx1Out). In addition to this outside antenna, we installed two vertically polarized antennas at different locations inside the receiver vehicle as shown in Fig. \ref{antenna}. 

\subsection{Measurement Settings }

We selected for each vehicular communication scenario a suitable and beneficial setup of measurement timing. One measurement run of the HHI channel sounder contains 10,000 snapshots. Instead of using all snapshots consecutively for a single continuous recording, we organize the available snapshots into sets of 6-13 snapshots. The time interval between the snapshots is 0.2-0.7~ms, which results in a set recording time of 2.4-3.6~ms. The time interval between the first snapshot of two consecutive sets is 10-100~ms. We recorded 769-1666~sets, which amounts to a total measurement time of 16-124~s per run. This measurement setup with gaps between recording sets permits longer measurement runs and in addition reflects the packet on-air time of cooperative awareness messages based on IEEE~802.11p. The length of the set recording time accounts for the maximum IEEE 802.11p frame duration of 2~ms for the maximum allowed payload of 1500~Bytes.

Summing up all 72 measurement runs, the performed analysis is based on 108~min of total measurement time. The total pure recoding time without gaps between sets amounts to 240~s. Since we conducted measurements with two transmitter (Tx) antennas and three receiver (Rx) antennas, the effective total measurement time increases by a factor six (antenna pairs) and adds up to a total of almost 11~h of measurement time or to a total of 24~min of pure recording time.

\subsection{Measurement Environments }

\begin{figure}[!t]
\centering
\includegraphics[width=2.7in]{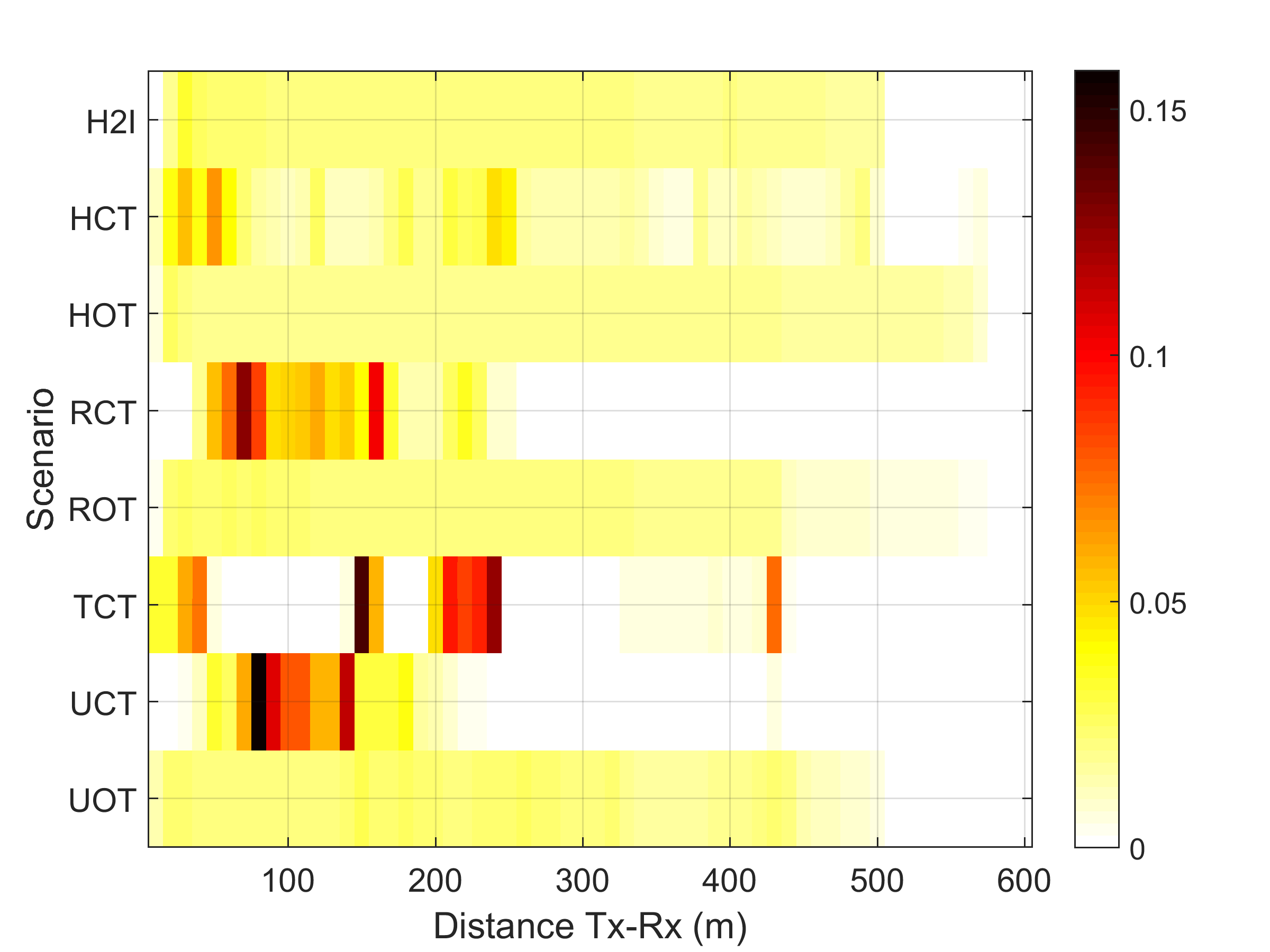}
\caption{Histogram of distance between measurement vehicles during channel recording sets.}
\label{MeasDist}
\end{figure}

\begin{figure}[!t]
\centering
\includegraphics[width=2.7in]{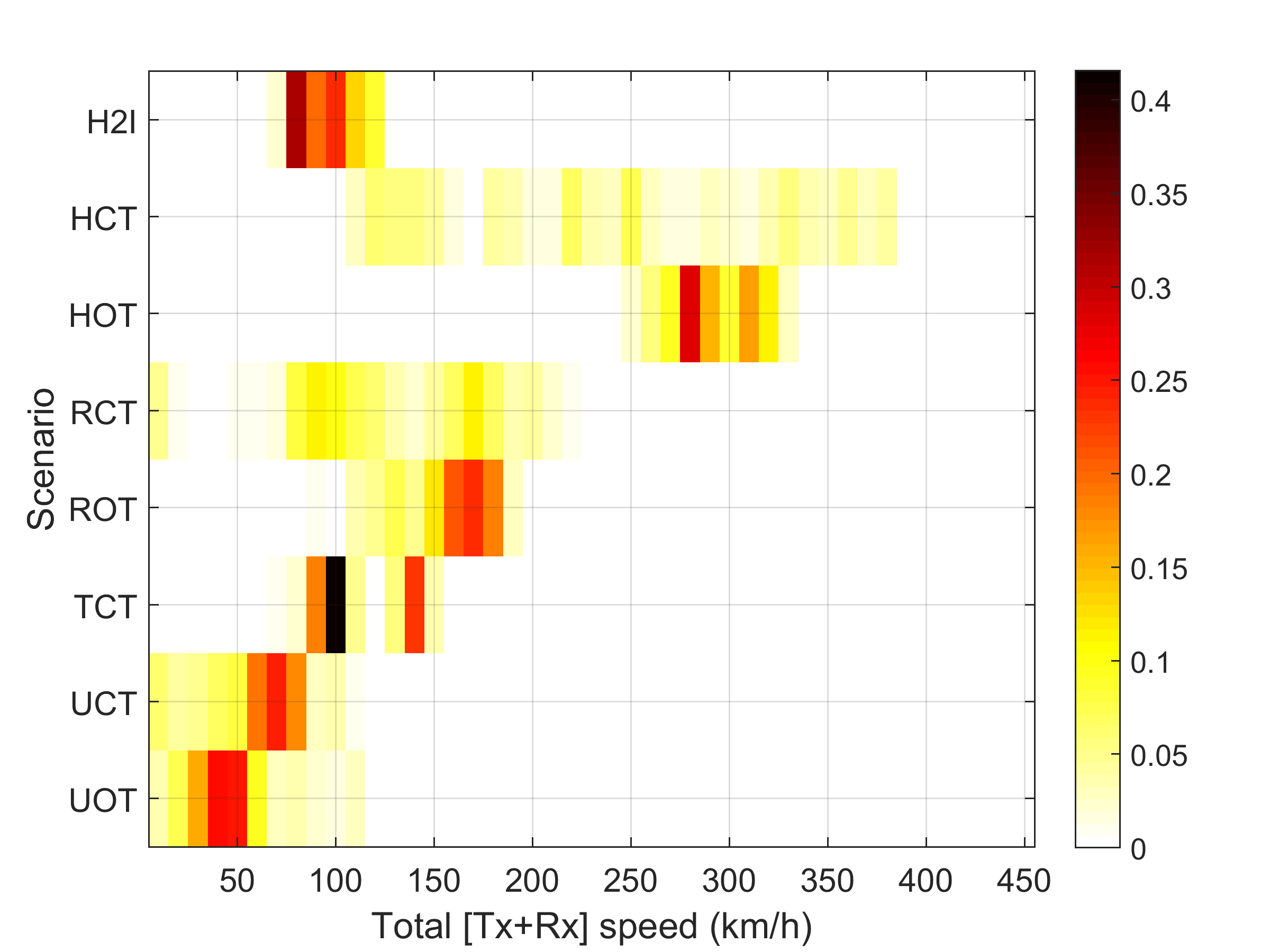}
\caption{Histogram of relative vehicle speed during channel recording sets.}
\label{MeasSpeed}
\end{figure}

We investigate eight relevant vehicular communication scenarios and selected corresponding measurement sites in or around Berlin, accordingly. During measurements, each vehicle was equipped with a video camera and a highly accurate Global Navigation Satellite System (GNSS) system for positioning. The histograms in Fig. \ref{MeasDist} give an overview of the distances between Tx and Rx during the measurement runs, whereas Fig. \ref{MeasSpeed} displays the corresponding relative (Tx+Rx) speed. The measured scenarios and corresponding sites are: 

\textbf{H2I} - Highway-to-Infrastructure: 12 measurement runs. The stationary infrastructure antenna was located near a former custom office called Zollamt Dreilinden. The antenna was mounted at 3.50~m height on the west side of highway, around 5~m away from the highway guardrail and 6~m from the custom office building. During the measurements, we observed a medium-range traffic density and several large vehicles. The distance between the Tx and Rx antennas is evenly distributed up to 500~m, with relative speeds between 80 and 120~km/h as indicated in Fig. \ref{MeasDist}  and \ref{MeasSpeed}.

\textbf{HCT} - Highway Convoy Traffic: 12 measurement runs. One measurement run was conducted northbound on the highway A100 between Spandauer Damm and Siemensdamm, with dense traffic and several large vehicles. The remaining 11~measurement runs were conducted clockwise on the highway A10, starting from the highway junction Kreuz Oranienburg to the junction Dreieck Spreeau. Some parts of the highway were under construction and therefore had a speed limit of 60~km/h. The traffic density was low to high range and included large vehicles. The distance between the measurement vehicles reached 500~m, but was mostly around 50~m or around 230~m, with relative (Tx+Rx) speeds between 100 and 400~km/h as indicated in Fig. \ref{MeasDist}  and \ref{MeasSpeed}.

\textbf{HOT} - Highway Oncoming Traffic: 10 measurement runs. The measurement vehicles drove in a medium range traffic density with many large vehicles on the highway A10 between the exit Birkenwerder and the exit Muehlenbeck. The distance between vehicles was evenly distributed up to 570~m, with relative speeds between 250 and 330~km/h, as indicated in Fig. \ref{MeasDist}  and \ref{MeasSpeed}. 

\textbf{RCT} - Rural Convoy Traffic: 7 measurement runs. The measurement vehicles drove on several roads west of Berlin: westbound on the road B5 between Staaken and Elstal (2~runs), southbound on the road L863 between Nauen and Ketzin (3~runs) and eastbound on the road L92/B273 between Paretz and Marquardt (2 runs). The traffic density was medium-range dense and during 4 of the measured runs, a large vehicle drove between the measurement vehicles and blocked the line-of-sight. The distance between the vehicles was mostly in the range 50-200~m, with speeds between 80 and 200~km/h, as indicated in Fig. \ref{MeasDist}  and \ref{MeasSpeed}.

\textbf{ROT} - Rural Oncoming Traffic: 7 measurement runs. The measurement vehicles drove in a low range traffic density on the road L20 between Gross Glienicke and Seeburg. One of the measurement vehicles was driving at a distance of 5-20~m behind one of these large obstructing vehicles: large goods vehicle, coaches, large caravan, site vehicles of type Mercedes-Benz Actros 4146 and a van of type Volkswagen Type 2. The distance between vehicles was evenly distributed up to 550~m, with relative speeds mostly between 150 and 180~km/h, as indicated in Fig. \ref{MeasDist} and \ref{MeasSpeed}. A detailed description of this scenario can be found in \cite{mahler15}.

\textbf{TCT} - Tunnel Convoy Traffic: 4 runs in 3 different tunnels. Two of the measurement runs took place southbound in the so-called Tiergartentunnel, one measurement run took place northbound on the highway A111 below Gorkistr. and one measurement run northbound on the highway A111 next to the crossing Ruppiner Chaussee / Im Waldwinkel. We observed a medium to high range traffic density with several large vehicles. The measurements were conducted at several distances with emphasis on the ranges below 50~m, 150-250~m and around 420~m with relative speeds between 80 and 150~km/h, as indicated in Fig. \ref{MeasDist}  and \ref{MeasSpeed}.

\textbf{UCT} - Urban Convoy Traffic: 8 runs at 5 different measurement sites. One run took place eastbound on Skalitzerstr. between Kottbusser Tor and Wrangelstr. with a high traffic density and few large vehicles, one run northbound on Warschauerstr. between Oberbaumbruecke and Gruenbergerstr. with high traffic density and few large vehicles, one run westbound on Otto-Braun-Str. between Alexanderstr. and Niederwallstr. with low to high range traffic density and few large vehicles, two runs westbound on Leipzigerstr. between Axel-Springer-Str. and Mauerstr. with low to high range traffic density in driving direction and high traffic density in opposite driving direction and three runs westbound Buelowstr./Kleiststr. between Zietenstr. and Passauerstr. with medium to high range traffic including numerous large buses. Note that some runs were conducted next to elevated railways with numerous large metallic pillars at the height of the measurement antennas, which result in a rich and dynamic propagation environment. The distance between vehicles was usually 50 and 200~m, with relative speeds mostly between 60 and 80~km/h, as indicated in Fig. \ref{MeasDist}  and \ref{MeasSpeed}.

\textbf{UOT} - Urban Oncoming Traffic: 12 runs at 7 different measurement sites. One run took place on Strasse des 17.~Juni between Ernst-Reuter-Platz and Kloppstr. with low to medium range traffic density and very few large vehicles, two runs on Knobelsdorffstr. between Sophie-Charlotte-Str. and Schlossstr. with low to medium range traffic density, two runs on Danckelmannstr. between Seelingstr. and Kaiserdamm with low traffic density and few large vehicles standing, two runs on Wilmersdorferstr. between Otto-Suhr-Allee and Bismarckstr. with low to medium range traffic density and several large vehicles standing, two runs on Schillerstr. between Leibnizstr. and Wilmersdorferstr. with low to medium range traffic density and few large vehicles, one run on Bleibtreustr. between Mommsenstr. and Pestalozzistr. with medium range traffic density and two runs on Hardenbergstr. between Joachimsthalerstr. and Ernst-Reuter-Platz with medium range traffic density and few large vehicles. The distance between vehicles was evenly distributed up to 500~m, with relative speeds between 5-100~km/h but mostly around 50~km/h, as indicated in Fig. \ref{MeasDist}  and \ref{MeasSpeed}.

\section{MPC Parameter Extraction}
\subsection{Extraction Methods}

Our processing of the channel sounder data is a multi-step process and starts with MPC detection, followed by short-term MPC tracking and finally long-term MPC tracking. A detailed explanation of the processing steps including tracking performance measures can be found in \cite{mahler16}. In this paper, we apply these extraction methods on a large set of channel data, without introducing any new extraction methods. The processing starts with the detection of MPCs in the channel impulse response at a certain time instance, basically a search and subtract method as in \cite{santos10}: 
\begin{itemize}
\item Estimate the noise floor level, add 6 dB to obtain a noise floor threshold and set all values in the impulse response below this threshold to zero.
\item Find the strongest peak and save this as a detected MPC.
\item Subtract the channel sounder pulse at the detected MPC delay position from the measured transfer function.
\item Repeat until no additional MPCs are detected.
\end{itemize}
The subsequent short-term MPC tracking is explained in detail in \cite{mahler16}, the steps can be summarized as follows:
\begin{itemize}
\item Start in the first snapshot with the strongest peak and search in the second snapshot for neighboring peaks.
\item Use the observed delay change and magnitude change to define a two-dimensional search range and predict the peak location in the third snapshot. If a peak is found, an MPC track has been identified.
\item Use the latest delay change and magnitude change to get the next search ranges accordingly. Continue searching peaks along the MPC track, until no peak within the current search ranges is found.
\end{itemize}
In order to investigate the large-scale evolution of MPC tracks, a supplementary long-term tracking algorithm is applied. This additional tracking method interrelates MPC tracks across adjacent recording sets and gaps of 10-100 ms, as described in Section II.B. For this algorithm, only full-lifetime MPCs are considered, i.e. MPC tracks with a lifetime equal to the duration of the recording set. 
Disregarding non-full-lifetime MPC tracks results in a power loss, but increases the reliability of the long-term tracking.
The long-term tracking algorithm is quite simple and starts with the strongest MPC track in the current set, defines a two-dimensional search range and searches in the next set for possible candidates. In the next step of the algorithm, the delay change (Doppler frequency) of the current MPC track is used to predict the delay location of the MPC track in the next set. The same is done in the opposite direction; the delay change of the MPC track in the next set is used to predict the delay location of the MPC track in the current set. 
The deviation between the actual delay value and the predicted delay value is compared to a threshold value  $\chi$, again for both directions. Two MPC tracks are found to be related, if both deviations are below this threshold.

The tracking parameters in \cite{mahler16} are matched to channel data from a TCT scenario with dense multipath inter-arrival times and a gap of 10 ms between adjacent recording sets. In the current paper, we analyze channel data from various scenarios, where usually less dense multipath inter-arrival times occur and the set period intervals are in the range 10-100 ms. In order to apply the tracking algorithms on diverse scenarios, we doubled the delay threshold to $\chi = 2$ ns. 

The applied short-term tracking method does not capture diffuse parts of the propagation channel, which leads to a certain power  loss, as explained in \cite{mahler16}. In addition, disregarding non-full-lifetime MPCs leads to a further power loss. For the channel data analyzed in this paper we observe an average power loss of 2.2 dB due to the detection method and the short-term tracking, with values ranging from 2.1 dB for ROT to 2.5 dB for UCT. The power loss due to the long-term tracking is on average 1.3 dB, ranging from 0.8 dB for TCT to 1.8 dB for RCT. The total power loss due to all processing steps is on average 3.5 dB, ranging from 2.9 dB for TCT to 3.9 dB for RCT. 

\subsection{Extracted Parameters}

For a description of the five extracted and analyzed channel parameters, we start with the time-variant channel impulse response obtained from a wideband channel sounder
\begin{equation}\label{gl:barwq}
h(t, \tau) = \sum^{P(t)}_{k=1} a_k(t) e^{j \varphi_k} w ( \tau - \tau_k(t)), 
\end{equation}
where $a_k$ is the amplitude, $\varphi_k$ the phase, $\tau_k$ the delay of MPC $k$ and $w(\tau)$ the isolated channel sounder pulse. $P(t)$ is the number of MPCs at time $t$. Our channel data is discretized and we therefore translate the continuous form $h(t, \tau)$ to $h(i T_r, u T_b) = h[i,u]$, where $T_r$ is the recording set period, $T_b = 1/B$ is the delay resolution period and $B$ the measurement bandwidth of the channel sounder. More explanations on this discretization can be found in \cite{mahler16}.

Now, similar to definitions in \cite{he15}, we define three sets of MPCs:
\begin{itemize}
\item $\mathcal{L}_{i}$ is the set of all MPCs that exist at time instance $i$ and $P[i]$ is the number of MPCs at time instance $i$.
\item $\mathcal{L}_{i \rightarrow i}$ is the set of MPCs that are firstly observed at time instance $i$, whose index of path is $k_{i \rightarrow i} = 1, 2, ... , P_b[i]$. Hence, $P_b[i]$ is the number of newly observed MPCs at time instance $i$.
\item $\mathcal{L}_{\Delta i}$ is the set of MPCs that are firstly observed during the time period $\Delta i = b-i$ and we designate $R[i]$ to be the total number of newly observed MPCs within this time period.
\end{itemize}
In order to obtain $R[i]$, we define
\begin{equation}\label{gl:barwq}
\Delta i \simeq \frac{\Delta s}{T_r(v_{Tx}+v_{Rx})},
\end{equation}
where $v_{Tx}+v_{Rx}$ is the current relative speed of the measurement vehicles and $\Delta s$ is the traveled distance, set to be 1 m. Note that in order to maintain consistency among the scenarios, we exclude all data where the speed of one the measurement vehicles is below 5 km/h. Without this exclusion, we would obtain very high counts of newborn MPCs (newly observed MPCs) in case the vehicles are very slow or standing. The total number of newborn MPCs within 1 m distance traveled is then found by
\begin{equation}\label{gl:birth}
R[i] = P_b[i-b] + P_b[i-b+1] + ... + P_b[i] = \sum^i_{c=i-b} P_b[c].
\end{equation}
The long-term tracking (43) in \cite{mahler16} yields
\begin{equation}\label{gl:barwq}
\boldsymbol{q}_k[i] = \left( \begin{array}{c}\bar a_k[i] \\ \bar \tau_k[i] \\ \nu_k[i]  
\end{array} \right), 
\end{equation}
where $\bar a_k[i]$ is the mean amplitude, $\bar \tau_k[i]$ the mean delay and $\nu_k[i]$ the Doppler frequency of path $k$ during the recording set at time instance $i$. Based on this result we define the set 
\begin{equation}\label{gl:barwq}
\boldsymbol{Q}_k = \{ \boldsymbol{q}_k[i]\}_{i=i_k^{start}}^{I_{k}},
\end{equation}
with $i_k^{start}$ as the time instance where path $k$ appears for the first time and $I_k$ being the last time instance of this path. The lifetime of path $k$ is $\Psi_k = I_k - i_k^{start}$. For a better geometrical relevancy, we translate the lifetime to distance with
\begin{equation}\label{gl:life}
Y_k = \Psi_k T_r (v_{Tx}+v_{Rx}).
\end{equation}
The mean delay of a newborn MPC is $\bar \tau_b[i]$ with $b=1,2,..., P_b[i]$ and the excess delay of newborn MPCs therefore   
\begin{equation}\label{gl:delay}
\tau^x_b[i] = \bar \tau_b[i] - \tau_{LOS}[i],
\end{equation}
where $\tau_{LOS}[i]$ is the delay of the line-of-sight path at time $i$.

The Doppler frequency $\nu_k$ is based on the delay change and estimated for each MPC track, as described by (30) and (31) in \cite{mahler16}. Based on the Doppler frequency of newborn MPCs $\nu_b[i]$ with $b=1,2,..., P_b[i]$, we define 
\begin{equation}\label{gl:doppler}
\nu^n_b[i]= \frac{\nu_b[i]}{\frac{f_c}{c_o}(v_{Tx}+v_{Rx})}
\end{equation}
as the relative Doppler frequency of newborn MPCs, where $f_c$ is the carrier frequency and $c_o$ is the speed of light. The relativization accounts for different speeds of the measurement vehicles and makes it feasible to merge measurement data with different speeds for an ensemble characterization per scenario. Again, we exclude data where the speed of any measurement vehicle drops below 5 km/h.

\section{Results}


\subsection{Number of MPCs }
\begin{figure}[!t]
	\centering
	\includegraphics[width=2.5in]{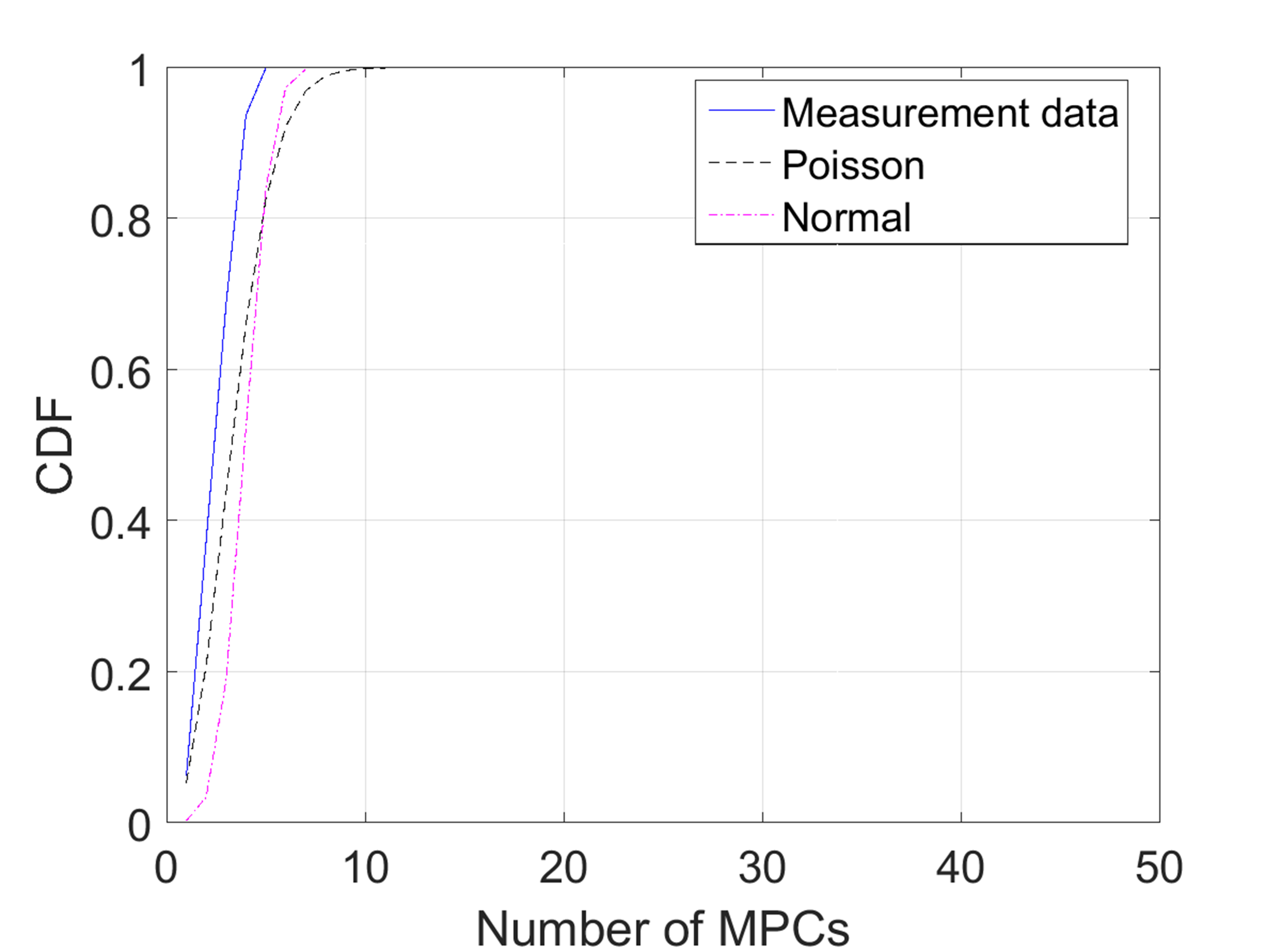}
	\caption{Comparison of number of MPCs distributions for an UOT measurement run at Tx-Rx distances of 450 m.}
	\label{CompareNumMPC_a}
\end{figure}
\begin{figure}[!t]
	\centering
	\includegraphics[width=2.5in]{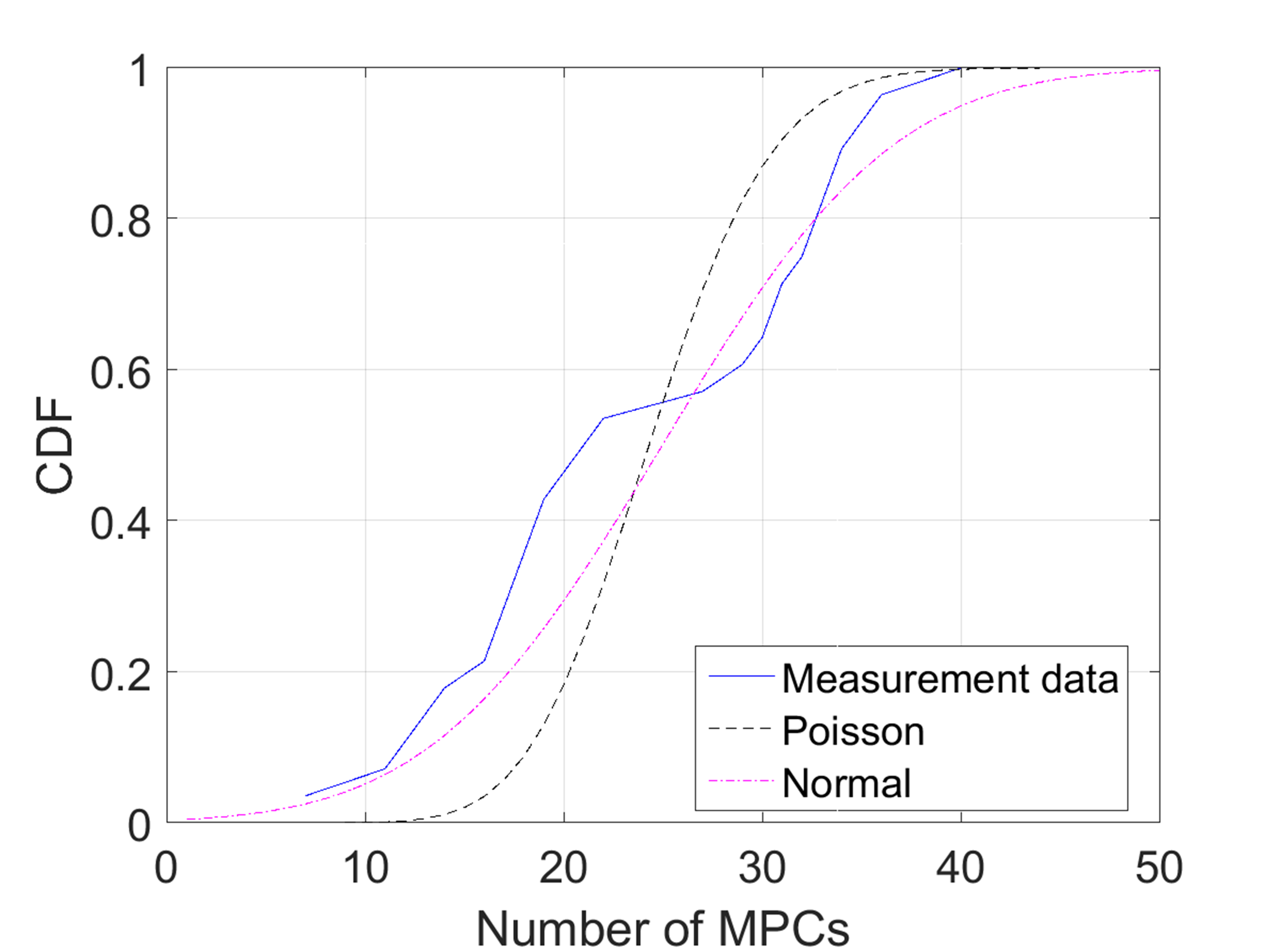}
	\caption{Comparison of number of MPCs distributions for an UOT measurement run at Tx-Rx distances of 80 m.}
	\label{CompareNumMPC_b}
\end{figure}
\begin{figure}[!t]
	\centering
	\includegraphics[width=2.5in]{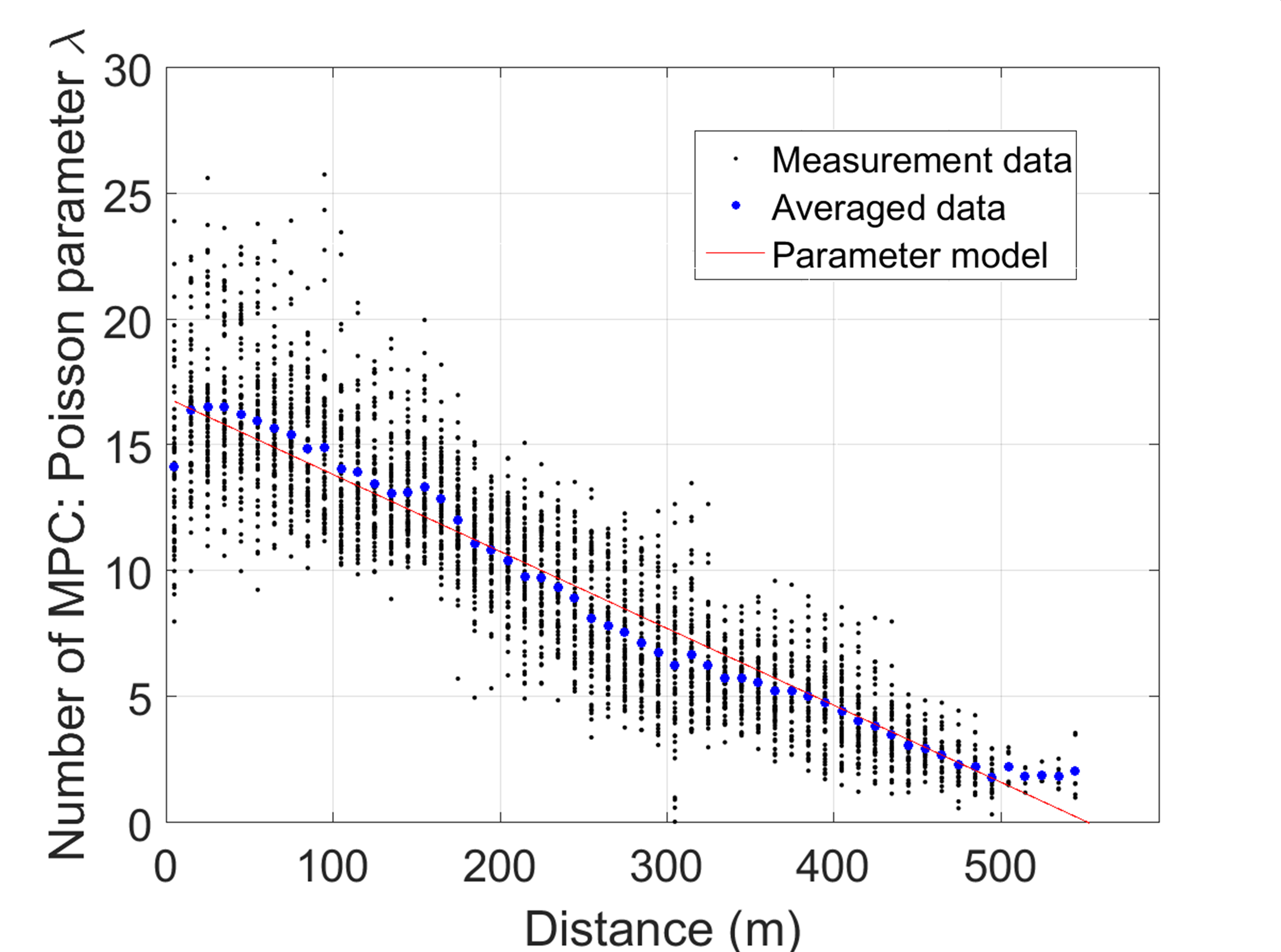}
	\caption{Distance dependence of the Poisson parameter $\lambda$ for the number of MPCs in the UOT scenario.}
	\label{NumMPCLambda_a}
\end{figure}
\begin{figure}[!t]
	\centering
	\includegraphics[width=2.5in]{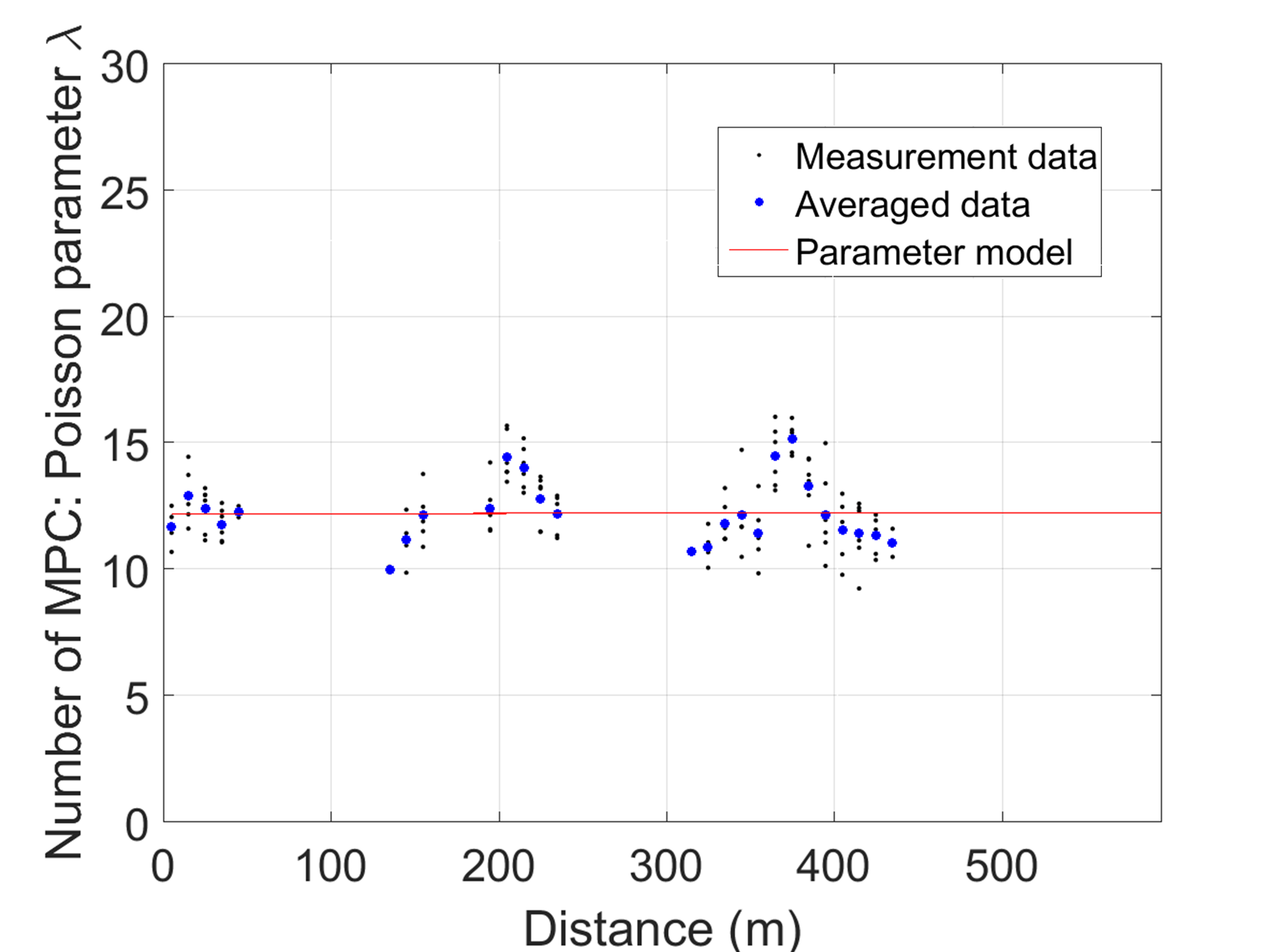}
	\caption{Distance dependence of the Poisson parameter $\lambda$ for the number of MPCs in the TCT scenario.}
	\label{NumMPCLambda_b}
\end{figure}
\begin{table}[]
	\centering
	\caption{Number of MPCs parameters.}
	\label{tab:num}
	\begin{tabular}{@{}llllll@{}}
		\toprule
		Scenario & STDEV & $p_0$ & $p_1$               & $p_2$              & MSE   \\ \midrule
		H2I      & 2.08  & 18.9  & -5.37$\cdot10^{-2}$ & 3.99               & 1.101 \\
		HCT      & 1.63  & 10.2  & -1.62$\cdot10^{-2}$ & 0                  & 0.587 \\
		HOT      & 0.92  & 7.97  & -2.14$\cdot10^{-2}$ & 1.54$\cdot10^{-5}$ & 0.197 \\
		RCT      & 1.38  & 9.73  & -1.70$\cdot10^{-2}$ & 0                  & 0.443 \\
		ROT      & 1.62  & 7.11  & -0.95$\cdot10^{-2}$ & 0                  & 0.196 \\
		TCT      & 0.87  & 12.2  & 0                   & 0                  & 1.612 \\
		UCT      & 1.69  & 14.5  & -2.61$\cdot10^{-2}$ & 0                  & 6.281 \\
		UOT      & 1.82  & 16.9  & -3.05$\cdot10^{-2}$ & 0                  & 0.604 \\ \bottomrule
	\end{tabular}
\end{table}
The number of MPCs $P[i]$ is an essential channel parameter and often modeled with the Poisson distribution
\begin{equation}\label{gl:poisson}
f(x|\lambda) = \frac{\lambda^x}{x!}e^{-\lambda}, x=0,1,2,...,\infty , 
\end{equation}
where the single parameter $\lambda$ influences both the mean and the variance of the distribution curve. We fit the cumulative density function (CDF) of the Poisson distribution to the measured data, together with a discretized Normal distribution fit for comparison. Both plots in Fig. \ref{CompareNumMPC_a} and \ref{CompareNumMPC_b} are from the same measurement run in an UOT scenario: plot (a) was measured at a Tx-Rx distance of 450~m, whereas plot (b) was measured at a distance of 80 m. In order to grasp this apparent distance dependency, we grouped the data into distance bins of 10~m width, computed the mean and the variance per bin and fitted a polynomial function
\begin{equation}\label{gl:lambda}
\lambda(d) = p_0 + p_1 d + p_2 d^2 , 
\end{equation}
with $d$ being the distance between Tx and Rx in meters. Note that $\lambda(d)$ becomes a line if $p_2$ is set to zero. As can be found in Fig. \ref{NumMPCLambda_a}, the UOT scenario exhibits a strong distance dependency. The mean standard deviation (STDEV) for this scenario is 1.82 and the mean square error (MSE) of the fitted line to the mean values is on average 0.604, as listed in Table \ref{tab:num}. Most other scenarios show the same behavior with similar values and can be modeled with a linear characteristic of $\lambda(d)$. The strongest parabolic characteristics was found in the H2I scenario with a $p_2$ value of around 4. A rather exceptional behavior is found for scenario TCT in Fig. \ref{NumMPCLambda_b}, where the waveguiding effect of tunnels lead to a distance-\textit{in}dependent Poisson parameter. The exceptionally high MSE values of the UCT scenario are due to high variances at distances below 200~m. 

We conducted the goodness-of-fit $\chi^2$-test and an MSE estimation for both distribution functions. Due to the distance dependency, this was done per 10~m distance bin. The $\chi^2$-tests result in a null hypothesis rejection rate of 13.74\% for the Poisson distribution and a rejection rate of 20.60\% for the Normal distribution (5\% significance level) . The average MSE$_N$ of the Poisson distribution fitting is $9.8\cdot10^{-3}$, ranging from $2.9\cdot10^{-3}$ for TCT to $13.9\cdot10^{-3}$ for HOT (see Table \ref{tab:mse}). The average MSE of the Normal distribution fit per distance bin is $1.41\cdot10^{-2}$, ranging from $0.31\cdot10^{-2}$ for TCT to $2.86\cdot10^{-2}$ for HOT. Hence, the Poisson distribution is a better choice both in terms of goodness-of-fit and accuracy.

The example in Fig. \ref{CompareNumMPC_b} has a high measurement data variance and visible inspection of the distribution fits could come to the conclusion that the Normal distribution is a better fit. However, visible inspection of other examples often lead to the opposite conclusion that the Poisson distribution is a better fit. Hence, a overall conclusion should be based on the results from the $\chi^2$-tests and the average MSE values.

\subsection{MPC Birth Rate }

The birth rate $R[i]$ defined in (\ref{gl:birth}) is the number of newborn MPCs per meter traveled and was found to have a similar distance dependency as $P[i]$. Consequently, we model $R[i]$ also with (\ref{gl:poisson}) and (\ref{gl:lambda}) and most scenarios exhibit a linear $\lambda(d)$ progression (see Table \ref{tab:birth}). Exceptional behavior is found again for the H2I scenario with a parabolic progression. The data for the TCT scenario is limited and the corresponding plot therefore sparse (not included in this paper due to limited space). Based on visual inspection, a constant value of approximately $\lambda = 7.5$ could also be appropriate for the TCT scenario, i.e. $p_0 = 7.5$ and $p_1 = p_2 = 0$. 

\begin{figure}[!t]
	\centering
	\includegraphics[width=2.4in]{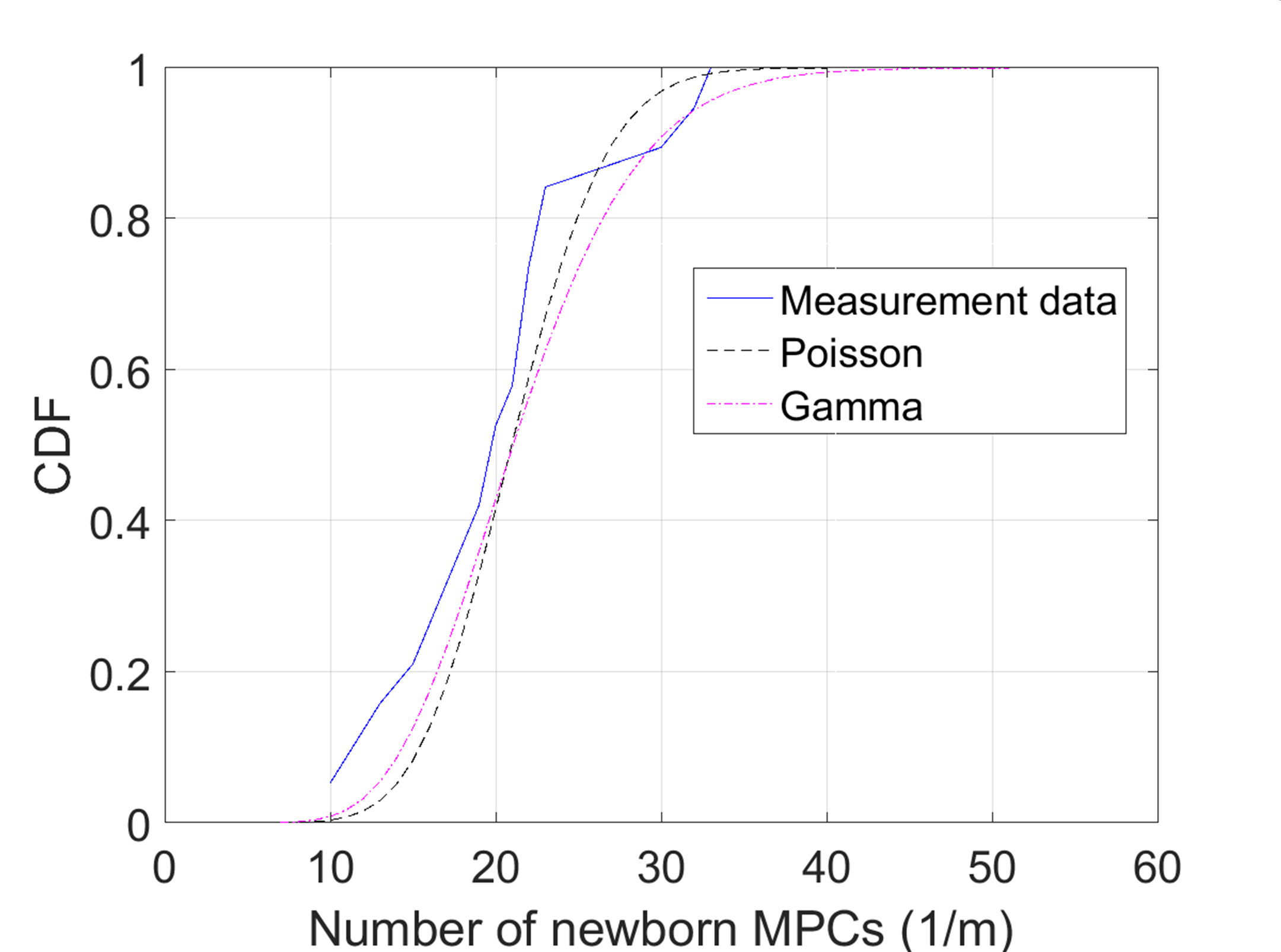}
	\caption{Comparison of birth rate distributions for an UOT measurement run at Tx-Rx distance of 290 m.}
	\label{CompareBirth_a}
\end{figure}
\begin{figure}[!t]
	\centering
	\includegraphics[width=2.4in]{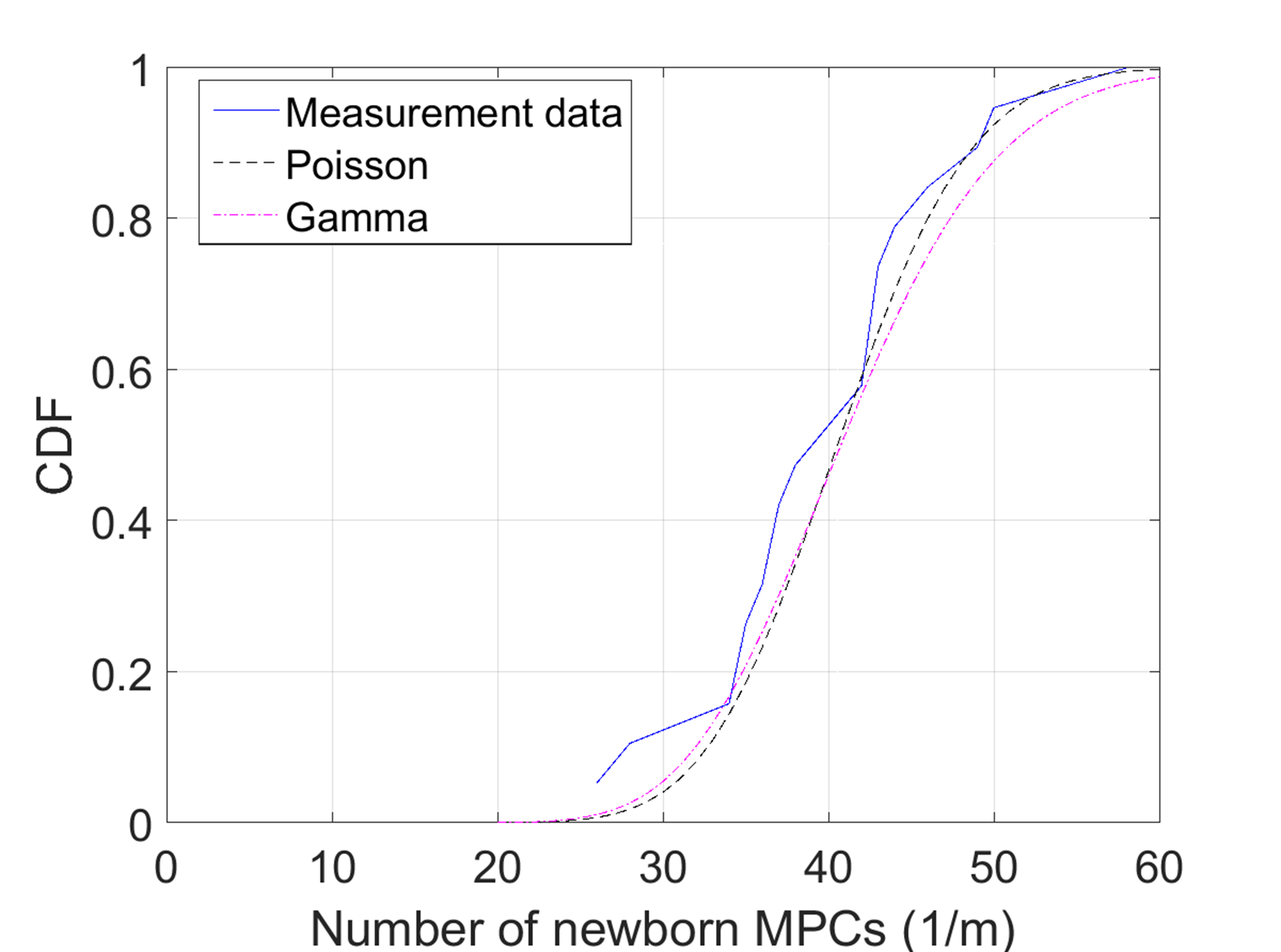}
	\caption{Comparison of birth rate distributions for an UOT measurement run at Tx-Rx distance of 270 m.}
	\label{CompareBirth_b}
\end{figure}
\begin{figure}[!t]
	\centering
	\includegraphics[width=2.4in]{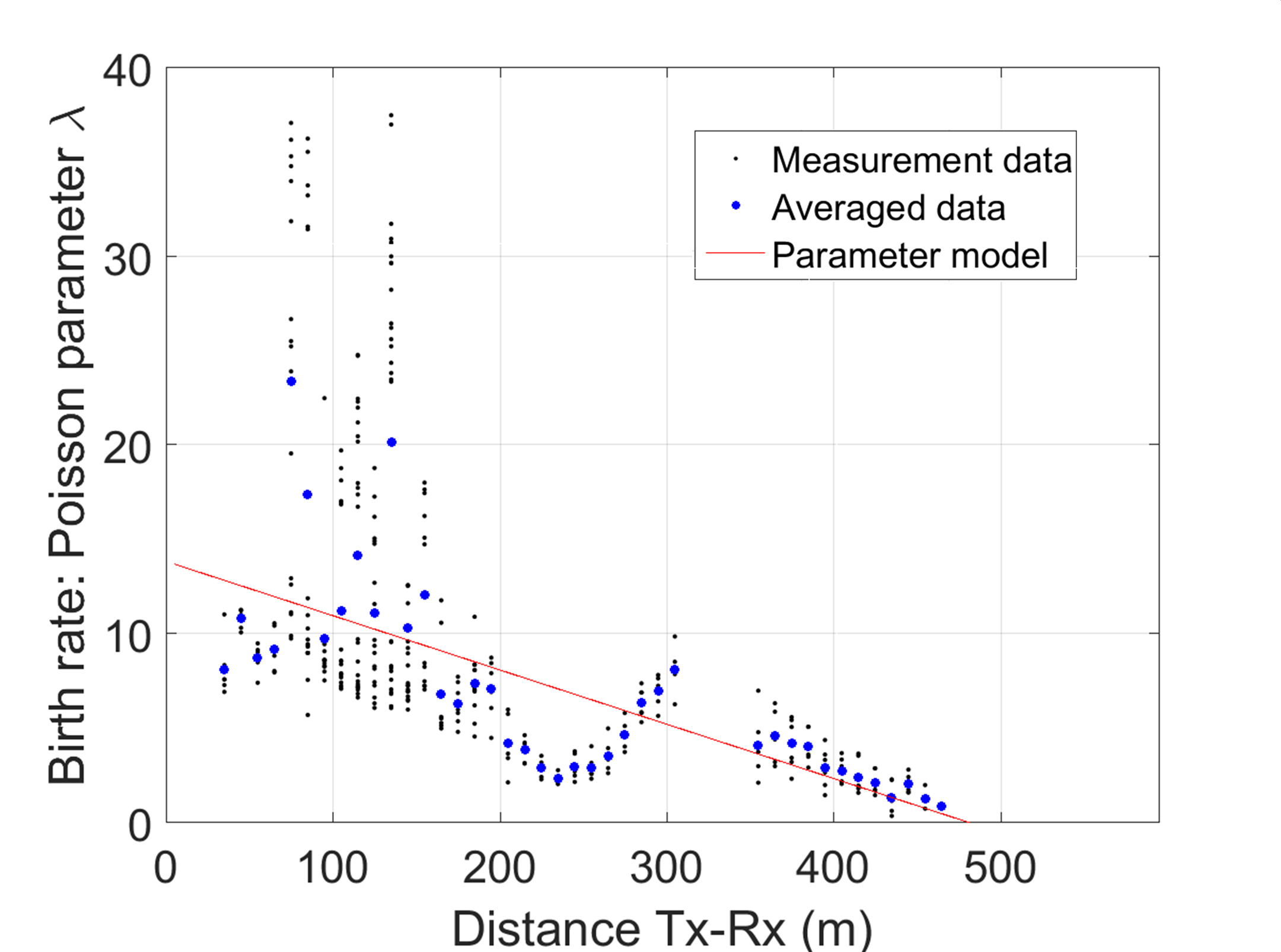}
	\caption{Distance-dependent progression of Poisson parameter $\lambda$ for birth rate distributions in UCT scenario.}
	\label{BirthRateLambda_a}
\end{figure}
\begin{figure}[!t]
	\centering
	\includegraphics[width=2.4in]{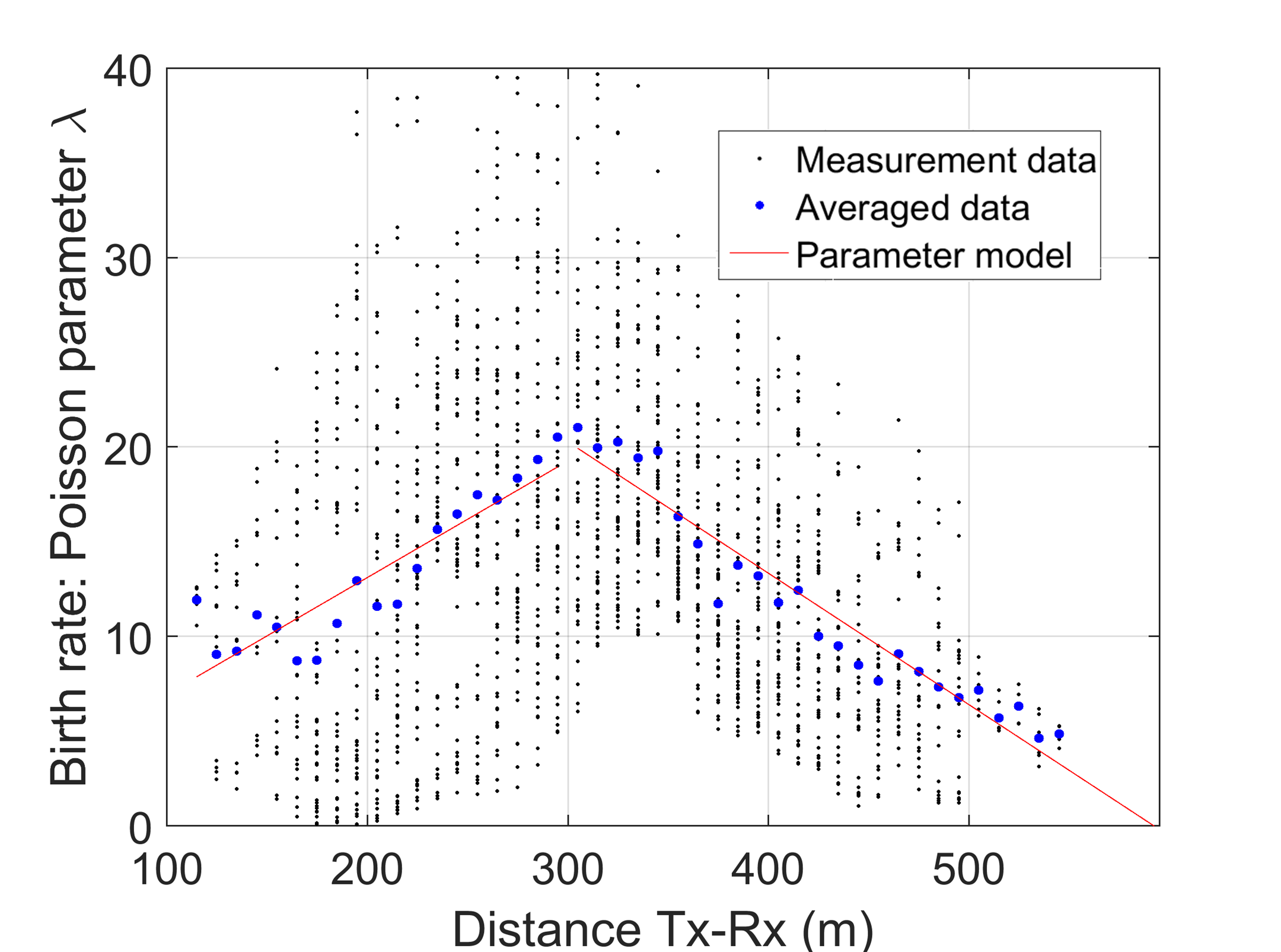}
	\caption{Distance-dependent progression of Poisson parameter $\lambda$ for birth rate distributions in UOT scenario.}
	\label{BirthRateLambda_b}
\end{figure}

The discretized Gamma distribution yields a good fit and was therefore included for comparison. The results from an UOT measurement run in Fig. \ref{CompareBirth_a} and \ref{CompareBirth_b} display a significant difference between the two empirical CDFs, which were recorded at distances of 290~m and 270~m. The example shows that the birth rate behavior in this scenario has a large variance and can change rapidly. We merge data from 12 runs at seven different measurement sites and see in Fig. \ref{BirthRateLambda_b} that the birth rate in the UOT scenario has a two-fold behavior. Similar to other scenarios, the progression above 300~m decreases with increased distance. However, the behavior is reversed for distances below 300~m. We compared each single UOT measurement run and observe this increasing-decreasing nature in all runs, however with varying widths and location of the maximum value. 

The $\chi^2$-tests and MSE estimations were again performed per 10~m distance bin. The $\chi^2$-tests result in a null hypothesis rejection rate of 8.82\% for the Poisson distribution and a rejection rate of 16.15\% for the Gamma distribution. The average MSE$_B$ of the Poisson distribution fitting is $7.2\cdot10^{-3}$, ranging from $1.4\cdot10^{-3}$ for TCT to $13.5\cdot10^{-3}$ for H2I (see Table \ref{tab:mse}). The average MSE of the Gamma distribution fitting per distance bin is $2.94\cdot10^{-2}$, ranging from $0.24\cdot10^{-2}$ for TCT to $11.9\cdot10^{-2}$ for HOT. Hence, the Poisson distribution is again a better choice both in terms of goodness-of-fit and accuracy. 

When comparing our results with \cite{he15}, we need to point out that our 1 GHz bandwidth resolves objects in the range of 0.3~m, whereas results in \cite{he15} are based on a 60 MHz measurement bandwidth, which leads to a delay resolution of 5~m. Hence, our derived results allow a far more detailed study of MPCs and are hardly comparable to \cite{he15}, since they are based on a different data set (measurement bandwidth) and a different analysis approach. The only parameter value derived both in our work and in \cite{he15} is the birth rate (Number of Newly Observed MPCs in \cite{he15}). However, no distance dependancy was found in \cite{he15} and a comparison is therefore not feasible or meaningful.

\begin{table}[]
\centering
\caption{Birth rate parameters.}
\label{tab:birth}
\begin{tabular}{@{}llllll@{}}
\toprule
Scenario & STDEV & $p_0$              & $p_1$               & $p_2$              & MSE    \\ \midrule
H2I      & 1.59  & 15.8               & -5.45$\cdot10^{-2}$ & 5.21               & 0.579  \\
HCT      & 1.18  & 7.26               & -1.25$\cdot10^{-2}$ & 0                  & 0.344  \\
HOT      & 0.58  & 5.26               & -1.77$\cdot10^{-2}$ & 1.58$\cdot10^{-5}$ & 0.0745 \\
RCT      & 1.06  & 6.90               & -1.25$\cdot10^{-2}$ & 0                  & 0.231  \\
ROT      & 1.22  & 3.49               & -0.34$\cdot10^{-2}$ & 0                  & 0.258  \\
TCT      & 0.61  & 9.82               & -0.61$\cdot10^{-2}$ & 0                  & 2.54   \\
UCT      & 2.40  & 13.8               & -2.86$\cdot10^{-2}$ & 0                  & 12.78  \\
UOTA     & 4.90  & 41.1               & -6.93$\cdot10^{-2}$ & 0                  & 2.004  \\
UOTB     & 8.24  & 9.01$\cdot10^{-1}$ & 6.13$\cdot10^{-2}$  & 0                  & 2.878  \\ \bottomrule
\end{tabular}
\end{table}

\subsection{MPC Lifetime}
\begin{figure}[!t]
	\centering
	\includegraphics[width=2.5in]{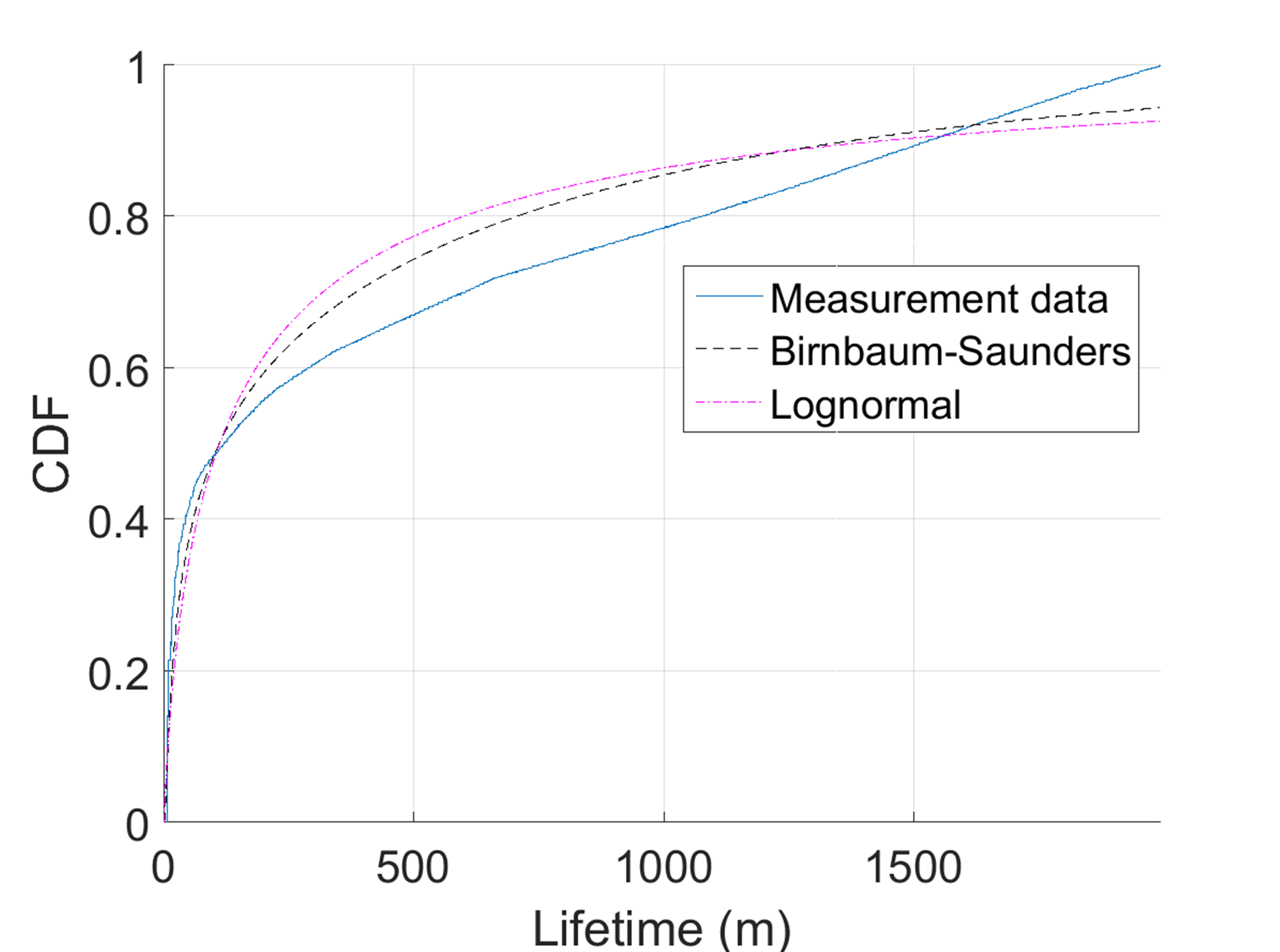}
	\caption{Comparison of lifetime distributions based on a single HCT measurement run.}
	\label{CompareLifetime_a}
\end{figure}
\begin{figure}[!t]
	\centering
	\includegraphics[width=2.5in]{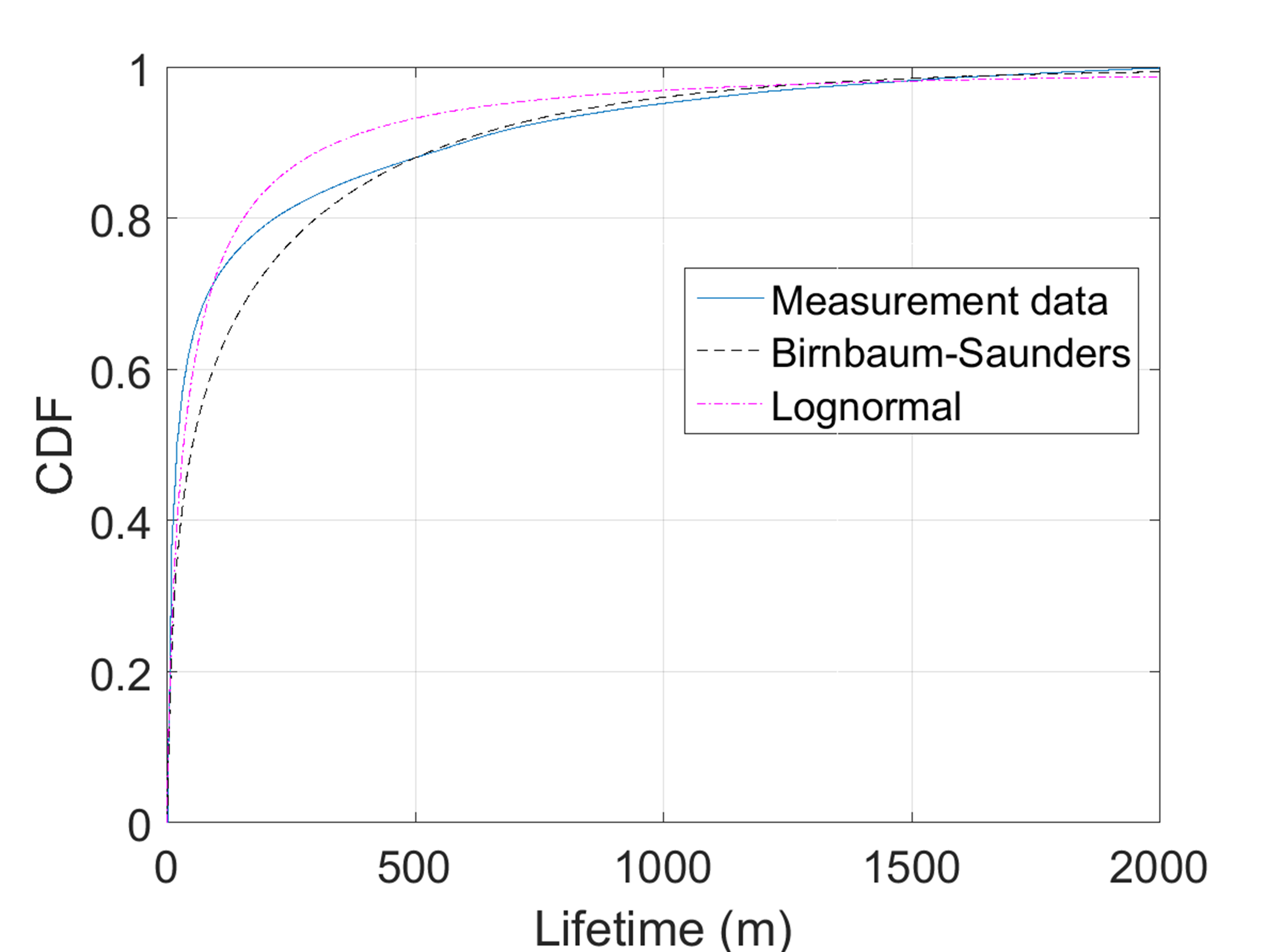}
	\caption{Comparison of lifetime distributions based on all data within HOT scenario.}
	\label{CompareLifetime_b}
\end{figure}
\begin{figure}
\centering
\includegraphics[width=3.2in]{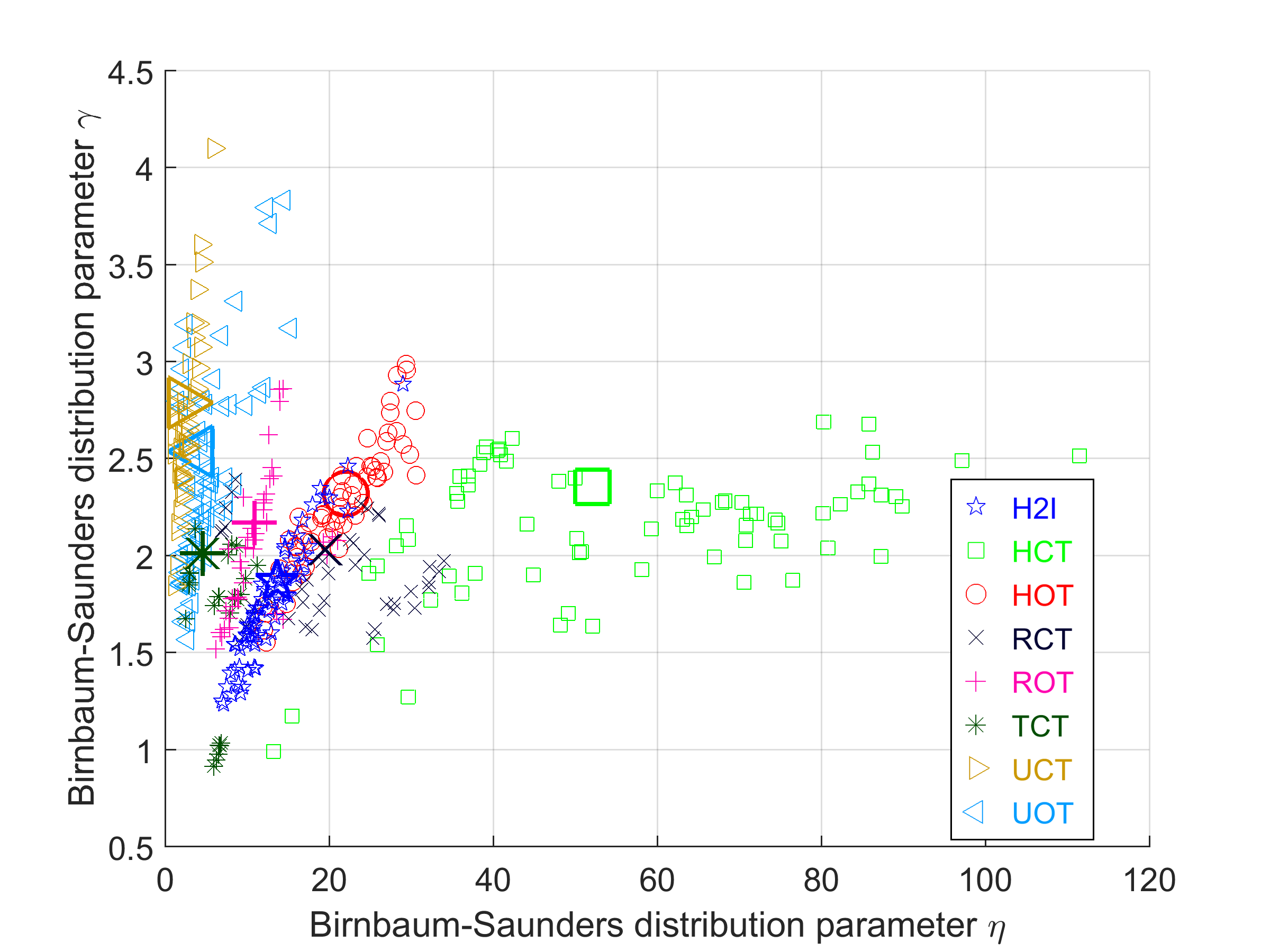}
\caption{Distribution of Birnbaum-Saunders parameters for lifetime occurrences per measurement (small marker) and per scenario (large marker).}
\label{LifetimeBirn}
\end{figure}

Finding a suitable distribution function for the MPC lifetime $Y_k$ defined in (\ref{gl:life}) appears to be problematic. We did not observe a distance dependency or any significant indications on a mixture distribution, i.e. a separation of shorter lifetime from longer lifetime MPCs with different thresholds did not improve the results. We fitted the Gamma, log-normal, Weibull and Inverse Gaussian distribution, but finally selected the Birnbaum-Saunders distribution
\begin{equation}\label{gl:barwq}
\begin{split}
f(x|\eta, \gamma) = \frac{1}{\sqrt{2 \pi}}  \left(  \frac{\left( \sqrt{^x/_\eta} - \sqrt{^\eta/_x}  \right)}{2\gamma x}   \right)   \\ \cdot   \exp \left\{-\frac{\left( \sqrt{^x/_\eta} - \sqrt{^\eta/_x}  \right)^2}{2\gamma^2} \right\}  , x>0  
\end{split}
\end{equation}
as the best choice, where $\eta$ affects the scale and $\gamma$ the shape of the curve. We identify the log-normal distribution as the best alternative and compare the fitting results. Fig. \ref{CompareLifetime_a} shows the \textit{worst} encountered fitting results for an HCT measurement run, where the fitting errors are predominantly found for larger values. The fitting result improves if applied on an entire scenario data set, as shown for the HOT in Fig. \ref{CompareLifetime_b}. Nevertheless, the fitted distribution still deviates significantly from the empirical curve, whereas it captures the essential behavior. Consequently, the $\chi^2$-tests yield a rejection rate of 99.31\% for the Birnbaum-Saunders distribution and a rejection rate of 100\% for the log-normal distribution. The average MSE$_L$ of the Birnbaum-Saunders distribution fitting per measurement is $2.0\cdot10^{-3}$, ranging from $0.95\cdot10^{-3}$ for TCT to $2.7\cdot10^{-3}$ for H2I (see Table \ref{tab:mse}). The average MSE of the Lognormal distribution fitting per measurement is $1.5\cdot10^{-3}$, ranging from $0.5\cdot10^{-3}$ for TCT to $2.8\cdot10^{-3}$ for HOT. 
The average MSE of the Birnbaum-Saunders distribution fitting per scenario is $0.87\cdot10^{-3}$, ranging from $0.40\cdot10^{-3}$ for TCT to $1.4\cdot10^{-3}$ for HCT. The average MSE of the Lognormal distribution fitting per scenario is $5.96\cdot10^{-4}$, ranging from $0.77\cdot10^{-4}$ for TCT to $16\cdot10^{-4}$ for HCT. The distribution of all estimated Birnbaum-Saunders parameters is displayed in Fig. \ref{LifetimeBirn}, where large markers indicate the parameters of the overall fit per scenario. We observe the shortest lifetime in urban or tunnel scenarios and the longest lifetime convoy traffic scenarios, in particular the HCT scenario. The parameter values for all scenarios can be found in Table \ref{tab:para}.

\subsection{Excess Delay of Newborn MPCs}

The excess delay $\tau^x_b[i]$, defined in (\ref{gl:delay}), was found to be non-distance-dependent and is modeled with the log-normal distribution

\begin{equation}\label{gl:barwq}
f(x|\psi, \rho) = \frac{1}{x \rho\sqrt{2 \pi}}\exp \left\{\frac{\left( \ln x - \psi \right)^2}{2\rho^2} \right\}    , x>0. 
\end{equation}
The parameter $\psi$ defines the peak location, whereas the parameter $\rho$ scales the curve. This distribution function yields a comparatively good fit, except for the scenario H2I. The large MSE value of 32$\cdot10^{-4}$ in Table \ref{tab:mse} is due to the fact that most runs in this scenario revealed two cores of excess delay occurrences. While most observed delay values are close to 0 ns, 5-30\% of the delays occurred at values of 300-350 ns. This is probably due to scattering at the buildings surrounding the infrastructure antenna and could be modeled with a mixture distribution. Since this issue of near and distant delay occurrences was only observed in the H2I scenario and a few urban measurement runs, we decided for simpler two-parameter distribution functions for the modeling. 

In addition to the log-normal distribution, we fitted an exponential distribution and found that the $\chi^2$-tests per measurement run result in a null hypothesis rejection rate of 80.56\% for the log-normal distribution and a rejection rate of 79.17\% for the exponential distribution. We also conducted a Kolmogorov-Smirnov test (KS) and found that the rejection rates of the log-normal and exponential distributions are  87.27\% and 98.38\%, respectively. The average MSE$_E$ of the log-normal distribution fitting per measurement is $1.1\cdot10^{-3}$, ranging from $0.36\cdot10^{-3}$ for UOT to $32\cdot10^{-3}$ for H2I (see Table \ref{tab:mse}). The average MSE of the exponential distribution fitting per measurement is $5.3\cdot10^{-3}$, ranging from $2.2\cdot10^{-3}$ for UCT to $10.7\cdot10^{-2}$ for H2I. The average MSE of the log-normal distribution fitting per scenario is $7.05\cdot10^{-4}$, ranging from $0.73\cdot10^{-4}$ for UCT to $32\cdot10^{-4}$ for H2I. The average MSE of the exponential distribution fitting per scenario is $4.2\cdot10^{-3}$, ranging from $1.0\cdot10^{-3}$ for UCT to $10.4\cdot10^{-3}$ for H2I.

\begin{figure}
\centering
\includegraphics[width=3.2in]{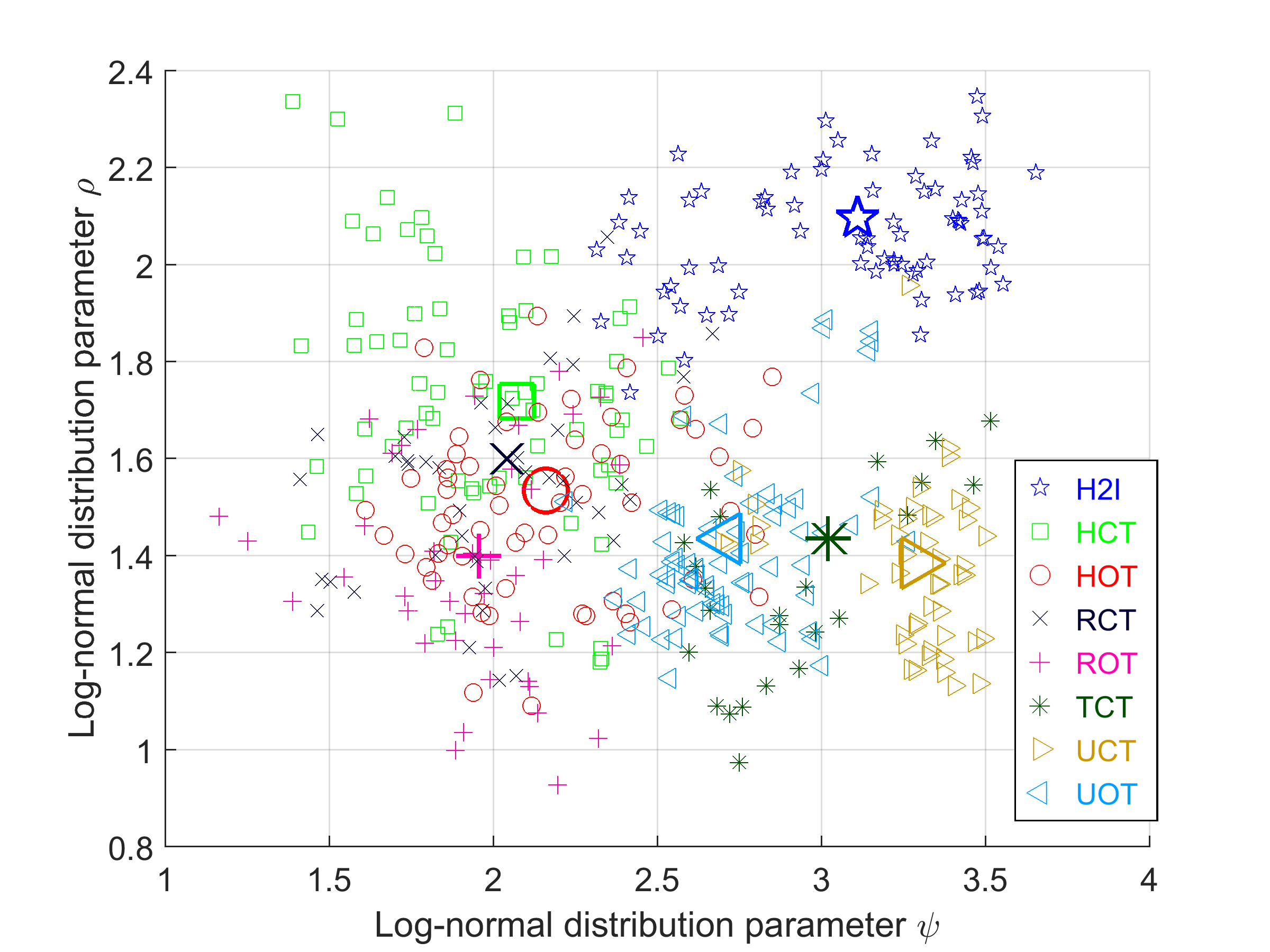}
\caption{Distribution of log-normal parameters for excess delay occurrences per measurement (small marker) and per scenario (large marker).}
\label{Delay}
\end{figure}

The goodness-of-fit test rates suggest slightly an advantage for the log-normal distribution, but essentially suggest a rejection of both distribution hypothesis. The exponential distribution is better in terms of convergence of the CDF towards 1 within the observed value range, which prevents the model from generating undesired very long excess delays. The fitting per measurement is more accurate with log-normal distributions, whereas the fitting per scenario is a little more accurate with exponential distributions, mainly due to better fits in the H2I scenario. The selection of log-normal is based on a better overall accuracy and slightly better goodness-of-fit test results. The distribution of the observed log-normal parameters in Fig. \ref{Delay} indicate larger excess delays in the urban scenarios and smaller excess delays in rural scenarios (see also Table \ref{tab:para}).

\subsection{Relative Doppler frequency of Newborn MPCs}

\begin{figure}
	\centering
	\includegraphics[width=2.5in]{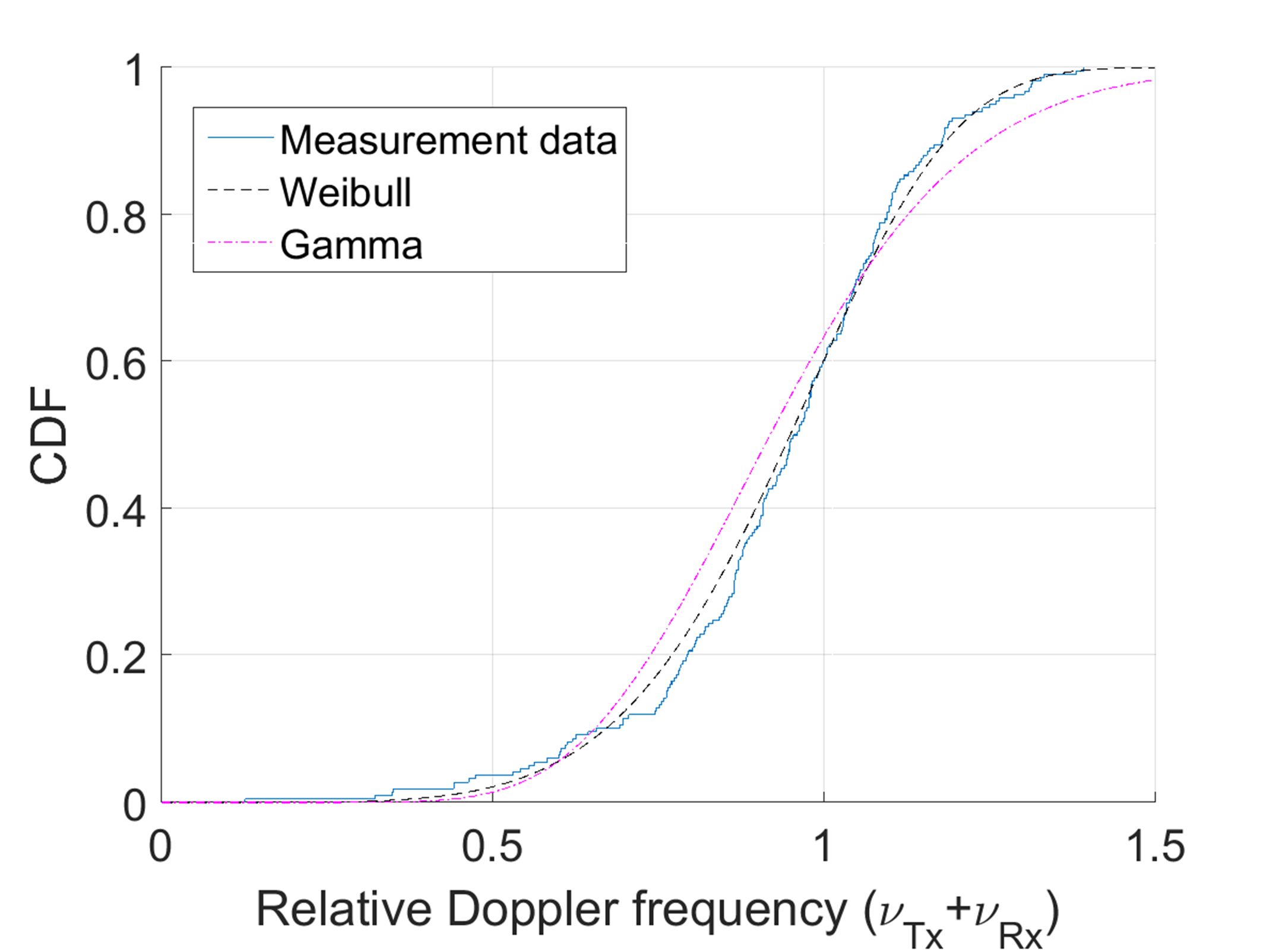}
	\caption{Comparison of relative Doppler distributions for measurement runs in HOT scenario.}
	\label{CompareDoppler_a}
\end{figure}

\begin{figure}
	\centering
	\includegraphics[width=2.5in]{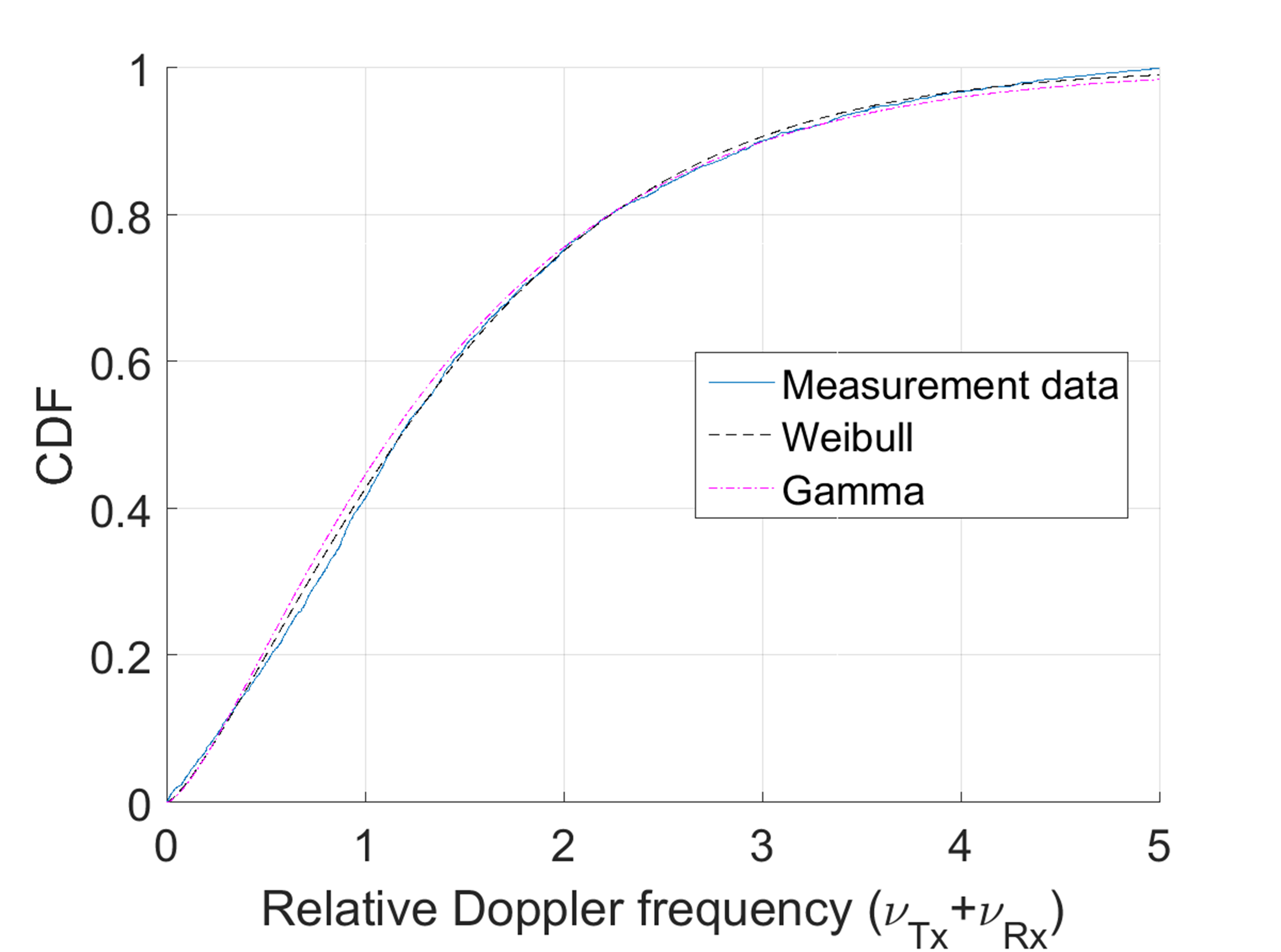}
	\caption{Comparison of relative Doppler distributions for measurement runs in UOT scenario.}
	\label{CompareDoppler_b}
\end{figure}

\begin{figure}
	\centering
	\includegraphics[width=3.3in]{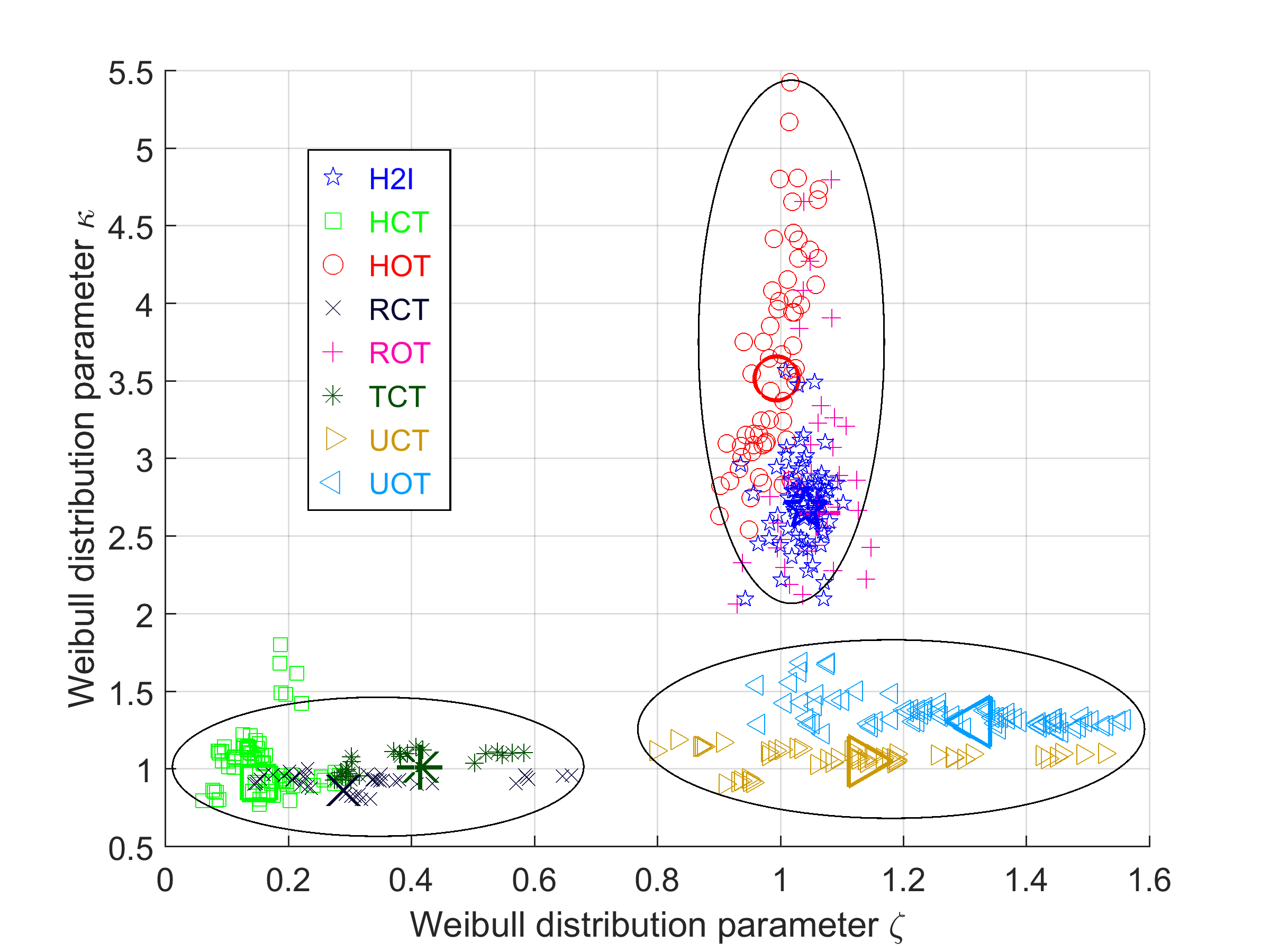}
	\caption{Distribution of Weibull parameters for relative Doppler occurrences per measurement (small marker) and per scenario (large marker).}
	\label{Doppler}
\end{figure}

The relative Doppler frequency of newborn MPCs $\nu^n_b[i]$ defined in (\ref{gl:doppler}) is modeled with the Weibull distribution

\begin{equation}\label{gl:barwq}
f(x|\zeta, \kappa) = \frac{\kappa}{\zeta} \left( \frac{x}{\zeta}  \right) ^{\kappa -1} \exp \left\{- \left( \frac{x}{ \zeta}  \right)^{\kappa} \right\}    , x>0. 
\end{equation}
The parameter $\zeta$ determines the scale and the parameter $\kappa$ the shape of the curve. For the HOT measurement run in Fig. \ref{CompareDoppler_a}, the Weibull distribution has a slightly better fit than the alternative Gamma distribution. Both distributions fit very well in an UOT scenario, as can be seen in Fig. \ref{CompareDoppler_b}. The measurement runs in Fig. \ref{CompareDoppler_a} and \ref{CompareDoppler_b} were selected as examples for fitting results, because they show that the Weibull distribution is suitable even for extreme values found within the parameter distribution in Fig. \ref{Doppler}, i.e. the HOT run is fitted with $\zeta=1.01$ and $\kappa=5.42$, whereas UOT is fitted with $\zeta=1.55$ and $\kappa=1.31$.

The $\chi^2$-tests result in a null hypothesis rejection rate of 91.67\% for the Weibull distribution and a rejection rate of 86.34\% for the Gamma distribution. The KS-tests revealed rejection rates of 62.50\% and 76.62\% for the Weibull and Gamma distributions respectively. The average Weibull MSE$_D$ is $8.2\cdot10^{-4}$, ranging from $0.95\cdot10^{-4}$ for UCT to $20\cdot10^{-4}$ for HOT (see Table \ref{tab:mse}). The average Gamma MSE is $2.1\cdot10^{-3}$, ranging from $0.12\cdot10^{-3}$ for UCT to $7.2\cdot10^{-3}$ for HOT. The Weibull MSE per scenario is $5.80\cdot10^{-4}$, ranging from $0.03\cdot10^{-3}$ for UCT to $1.6\cdot10^{-3}$ for HOT. The Gamma MSE per scenario is $1.6\cdot10^{-3}$, ranging from $0.03\cdot10^{-3}$ for UCT and $6.2\cdot10^{-3}$ for HOT. 

The distribution of the Weibull parameters in Fig. \ref{Doppler} reveals a distinctive characteristic and encourages the definition of three groups (see Table \ref{tab:para}): 
\begin{enumerate}
\item the urban traffic group (UCT, UOT) in the lower right corner in Fig. \ref{Doppler} is characterized by a curved CDF-shape and relative Doppler frequency values of up to 5 (see Fig. \ref{CompareDoppler_b})
\item the convoy traffic group (HCT, RCT, TCT) in the lower left corner in Fig. \ref{Doppler} is also characterized by a curved CDF-shape, however with lower values up to 2
\item the oncoming traffic group (H2I, HOT, ROT) in the upper part in Fig. \ref{Doppler} is characterized by a s-shaped CDF curve around the value 1 (in Fig. \ref{CompareDoppler_a})
\end{enumerate}
\begin{table}[]
	\centering
	\caption{Parameters of lifetime, excess delay and relative Doppler.}
	\label{tab:para}
	\begin{tabular}{@{}lllllll@{}}
		\toprule
		Scenario & $\eta$ & $\gamma$ & $\psi$ & $\rho$ & $\zeta$ & $\kappa$ \\ \midrule
		H2I      & 13.62  & 1.867    & 3.110  & 2.096  & 1.041   & 2.679    \\
		HCT      & 52.07  & 1.355    & 2.071  & 1.718  & 0.152   & 0.910    \\
		HOT      & 22.01  & 2.319    & 2.160  & 1.534  & 0.993   & 3.517    \\
		RCT      & 19.52  & 2.031    & 2.041  & 1.600  & 0.289   & 0.862    \\
		ROT      & 10.87  & 2.169    & 1.955  & 1.399  & 1.061   & 2.650    \\
		TCT      & 4.554  & 2.011    & 3.021  & 1.435  & 0.414   & 1.013    \\
		UCT      & 2.248  & 2.789    & 3.288  & 1.385  & 1.134   & 1.051    \\
		UOT      & 3.999  & 2.539    & 2.709  & 1.435  & 1.317   & 1.306    \\ \bottomrule
	\end{tabular}
\end{table}
\begin{table}[]
	\centering
	\caption{Mean square error of distribution fits.}
	\label{tab:mse}
	\begin{tabular}{@{}llllll@{}}
		\toprule
		Scenario & MSE$_N$            & MSE$_B$            & MSE$_L$           & MSE$_E$           & MSE$_D$           \\ \midrule
		H2I      & 1.84$\cdot10^{-2}$ & 1.35$\cdot10^{-2}$ & 2.7$\cdot10^{-3}$ & 34$\cdot10^{-4}$  & 14$\cdot10^{-4}$  \\
		HCT      & 0.96$\cdot10^{-2}$ & 0.61$\cdot10^{-2}$ & 1.9$\cdot10^{-3}$ & 7.3$\cdot10^{-4}$ & 8.6$\cdot10^{-4}$ \\
		HOT      & 1.39$\cdot10^{-2}$ & 0.70$\cdot10^{-2}$ & 1.6$\cdot10^{-3}$ & 6.2$\cdot10^{-4}$ & 20$\cdot10^{-4}$  \\
		RCT      & 0.51$\cdot10^{-2}$ & 0.34$\cdot10^{-2}$ & 1.7$\cdot10^{-3}$ & 5.7$\cdot10^{-4}$ & 3.1$\cdot10^{-4}$ \\
		ROT      & 1.61$\cdot10^{-2}$ & 0.89$\cdot10^{-2}$ & 2.4$\cdot10^{-3}$ & 7.5$\cdot10^{-4}$ & 11$\cdot10^{-4}$  \\
		TCT      & 0.29$\cdot10^{-2}$ & 0.14$\cdot10^{-2}$ & 0.9$\cdot10^{-3}$ & 17$\cdot10^{-4}$  & 5.5$\cdot10^{-4}$ \\
		UCT      & 0.41$\cdot10^{-2}$ & 1.08$\cdot10^{-2}$ & 2.3$\cdot10^{-3}$ & 4.5$\cdot10^{-4}$ & 0.9$\cdot10^{-4}$ \\
		UOT      & 0.82$\cdot10^{-2}$ & 0.61$\cdot10^{-2}$ & 2.7$\cdot10^{-3}$ & 3.6$\cdot10^{-4}$ & 1.8$\cdot10^{-4}$ \\ \midrule
		Mean     & 0.98$\cdot10^{-2}$ & 0.72$\cdot10^{-2}$ & 2.0$\cdot10^{-3}$ & 11$\cdot10^{-4}$  & 8.1$\cdot10^{-4}$
	\end{tabular}
\end{table}

\section{Conclusions}
For accurate modeling of channel properties in vehicular communication scenarios it is important to understand the behavior of MPCs. In this paper we have modeled the number of MPCs and the birth rate of MPCs with a distant-dependent Poisson distribution and we found exceptional behavior for the tunnel and urban scenarios. The statistical characterization of MPC lifetime appears still unclear, while the distribution of excess delay and relative Doppler frequency of individual MPCs are modeled accurately. The obtained statistical distributions can be used for a parametrization of suitable channel models and lead ultimately to a more accurate reproduction in V2V channel simulations.

\bibliographystyle{IEEEtran}
\bibliography{IEEEabrv,literature}

\begin{thebibliography}{10}
\providecommand{\url}[1]{#1}
\csname url@samestyle\endcsname
\providecommand{\newblock}{\relax}
\providecommand{\bibinfo}[2]{#2}
\providecommand{\BIBentrySTDinterwordspacing}{\spaceskip=0pt\relax}
\providecommand{\BIBentryALTinterwordstretchfactor}{4}
\providecommand{\BIBentryALTinterwordspacing}{\spaceskip=\fontdimen2\font plus
\BIBentryALTinterwordstretchfactor\fontdimen3\font minus
  \fontdimen4\font\relax}
\providecommand{\BIBforeignlanguage}[2]{{%
\expandafter\ifx\csname l@#1\endcsname\relax
\typeout{** WARNING: IEEEtran.bst: No hyphenation pattern has been}%
\typeout{** loaded for the language `#1'. Using the pattern for}%
\typeout{** the default language instead.}%
\else
\language=\csname l@#1\endcsname
\fi
#2}}
\providecommand{\BIBdecl}{\relax}
\BIBdecl

\bibitem{renaudin10}
O.~Renaudin, V.~M. Kolmonen, P.~Vainikainen, and C.~Oestges, ``Non-stationary
  narrowband {MIMO} inter-vehicle channel characterization in the 5-{GHz}
  band,'' \emph{IEEE Transactions on Vehicular Technology}, vol.~59, no.~4, pp.
  2007--2015, May 2010.

\bibitem{karedal10}
J.~Karedal, F.~Tufvesson, T.~Abbas, O.~Klemp, A.~Paier, L.~Bernado, and A.~F.
  Molisch, ``Radio channel measurements at street intersections for
  vehicle-to-vehicle safety applications,'' in \emph{Vehicular Technology
  Conference (VTC 2010-Spring), 2010 IEEE 71st}, May 2010, pp. 1--5.

\bibitem{molisch05}
A.~F. Molisch, ``Ultrawideband propagation channels-theory, measurement, and
  modeling,'' \emph{IEEE Transactions on Vehicular Technology}, vol.~54, no.~5,
  pp. 1528--1545, Sept 2005.

\bibitem{Mat11}
D.~W. Matolak and Q.~Wu, ``Channel models for {V2V} communications: A
  comparison of different approaches,'' in \emph{Proceedings of the 5th
  European Conference on Antennas and Propagation (EUCAP)}, April 2011, pp.
  2891--2895.

\bibitem{Wang09}
C.~x.~Wang, X.~Cheng, and D.~I. Laurenson, ``Vehicle-to-vehicle channel
  modeling and measurements: recent advances and future challenges,''
  \emph{IEEE Communications Magazine}, vol.~47, no.~11, pp. 96--103, November
  2009.

\bibitem{Patz1}
A.~Chelli and M.~Patzold, ``A non-stationary {MIMO} vehicle-to-vehicle channel
  model derived from the geometrical street model,'' in \emph{Vehicular
  Technology Conference (VTC Fall), 2011 IEEE}, Sept 2011, pp. 1--6.

\bibitem{Patz2}
N.~Avazov and M.~Patzold, ``A novel wideband {MIMO} car-to-car channel model
  based on a geometrical semi-circular tunnel scattering model,'' \emph{IEEE
  Transactions on Vehicular Technology}, vol.~65, no.~3, pp. 1070--1082, March
  2016.

\bibitem{Zaj1}
A.~G. Zajic and G.~L. Stuber, ``Three-dimensional modeling and simulation of
  wideband {MIMO} mobile-to-mobile channels,'' \emph{IEEE Transactions on
  Wireless Communications}, vol.~8, no.~3, pp. 1260--1275, March 2009.

\bibitem{Zaj2}
A.~G. Zajic, G.~L. Stuber, T.~G. Pratt, and S.~T. Nguyen, ``Wideband {MIMO}
  mobile-to-mobile channels: Geometry-based statistical modeling with
  experimental verification,'' \emph{IEEE Transactions on Vehicular
  Technology}, vol.~58, no.~2, pp. 517--534, Feb 2009.

\bibitem{Cheng1}
X.~Cheng, C.~X. Wang, B.~Ai, and H.~Aggoune, ``Envelope level crossing rate and
  average fade duration of nonisotropic vehicle-to-vehicle {R}icean fading
  channels,'' \emph{IEEE Transactions on Intelligent Transportation Systems},
  vol.~15, no.~1, pp. 62--72, Feb 2014.

\bibitem{Cheng2}
X.~Cheng, Q.~Yao, M.~Wen, C.~X. Wang, L.~Y. Song, and B.~L. Jiao, ``Wideband
  channel modeling and intercarrier interference cancellation for
  vehicle-to-vehicle communication systems,'' \emph{IEEE Journal on Selected
  Areas in Communications}, vol.~31, no.~9, pp. 434--448, September 2013.

\bibitem{Patz3}
A.~Borhani, G.~L. Stuber, and M.~Patzold, ``A random trajectory approach for
  the development of non-stationary channel models capturing different scales
  of fading,'' \emph{IEEE Transactions on Vehicular Technology}, vol.~PP,
  no.~99, pp. 1--1, 2016.

\bibitem{Patz4}
M.~Patzold and C.~A. Gutierrez, ``The wigner distribution of sum-of-cissoids
  and sum-of-chirps processes for the modelling of stationary and
  non-stationary mobile channels,'' in \emph{2016 IEEE 83rd Vehicular
  Technology Conference (VTC Spring)}, May 2016, pp. 1--5.

\bibitem{Patz5}
A.~G. Zajić, ``Impact of moving scatterers on vehicle-to-vehicle narrow-band
  channel characteristics,'' \emph{IEEE Transactions on Vehicular Technology},
  vol.~63, no.~7, pp. 3094--3106, Sept 2014.

\bibitem{bernado14}
L.~Bernadó, T.~Zemen, F.~Tufvesson, A.~F. Molisch, and C.~F. Mecklenbräuker,
  ``Delay and {D}oppler spreads of nonstationary vehicular channels for
  safety-relevant scenarios,'' \emph{IEEE Transactions on Vehicular
  Technology}, vol.~63, no.~1, pp. 82--93, Jan 2014.

\bibitem{Sen08}
I.~Sen and D.~W. Matolak, ``Vehicle-vehicle channel models for the 5-{GH}z
  band,'' \emph{IEEE Transactions on Intelligent Transportation Systems},
  vol.~9, no.~2, pp. 235--245, June 2008.

\bibitem{Aco07}
G.~Acosta-Marum and M.~A. Ingram, ``Six time- and frequency- selective
  empirical channel models for vehicular wireless {LANs},'' \emph{IEEE
  Vehicular Technology Magazine}, vol.~2, no.~4, pp. 4--11, Dec 2007.

\bibitem{he15}
R.~He, O.~Renaudin, V.~M. Kolmonen, K.~Haneda, Z.~Zhong, B.~Ai, and C.~Oestges,
  ``A dynamic wideband directional channel model for vehicle-to-vehicle
  communications,'' \emph{IEEE Transactions on Industrial Electronics},
  vol.~62, no.~12, pp. 7870--7882, Dec 2015.

\bibitem{karedal09}
J.~Karedal, F.~Tufvesson, N.~Czink, A.~Paier, C.~Dumard, T.~Zemen, C.~F.
  Mecklenbrauker, and A.~F. Molisch, ``A geometry-based stochastic {MIMO} model
  for vehicle-to-vehicle communications,'' \emph{IEEE Transactions on Wireless
  Communications}, vol.~8, no.~7, pp. 3646--3657, July 2009.

\bibitem{Ming13}
M.~Gan, Z.~Xu, C.~F. Mecklenbrauker, and T.~Zemen, ``Cluster lifetime
  characterization for vehicular communication channels,'' in \emph{2015 9th
  European Conference on Antennas and Propagation (EuCAP)}, May 2015, pp. 1--5.

\bibitem{Xu15}
Z.~Xu, M.~Gan, and T.~Zemen, ``{Cluster-based non-stationary vehicular channel
  model},'' in \emph{Proc. Eur. Conf. Antennas Propag.}, Apr 2016.

\bibitem{paschalidis08}
P.~Paschalidis, M.~Wisotzki, A.~Kortke, W.~Keusgen, and M.~Peter, ``A wideband
  channel sounder for car-to-car radio channel measurements at 5.7 {GH}z and
  results for an urban scenario,'' in \emph{Vehicular Technology Conference,
  2008. VTC 2008-Fall. IEEE 68th}.\hskip 1em plus 0.5em minus 0.4em\relax IEEE,
  2008, pp. 1--5.

\bibitem{mahler15}
K.~Mahler, W.~Keusgen, F.~Tufvesson, T.~Zemen, and G.~Caire, ``Propagation
  channel in a rural overtaking scenario with large obstructing vehicles,'' in
  \emph{2016 IEEE 83rd Vehicular Technology Conference (VTC Spring)}, May 2016,
  pp. 1--5.

\bibitem{mahler16}
K.~Mahler, W.~Keusgen, F.~Tufvesson, T.~Zemen, , and G.~Caire, ``Tracking of
  wideband multipath components in a vehicular communication scenario,''
  \emph{IEEE Transactions on Vehicular Technology}, vol.~PP, no.~99, pp. 1--1,
  2016.

\bibitem{santos10}
T.~Santos, J.~Karedal, P.~Almers, F.~Tufvesson, and A.~F. Molisch, ``Modeling
  the ultra-wideband outdoor channel: Measurements and parameter extraction
  method,'' \emph{IEEE Transactions on Wireless Communications}, vol.~9, no.~1,
  pp. 282--290, January 2010.

\end{thebibliography}

\begin{IEEEbiography}[{\includegraphics[width=1in,height=1.25in,clip,keepaspectratio]{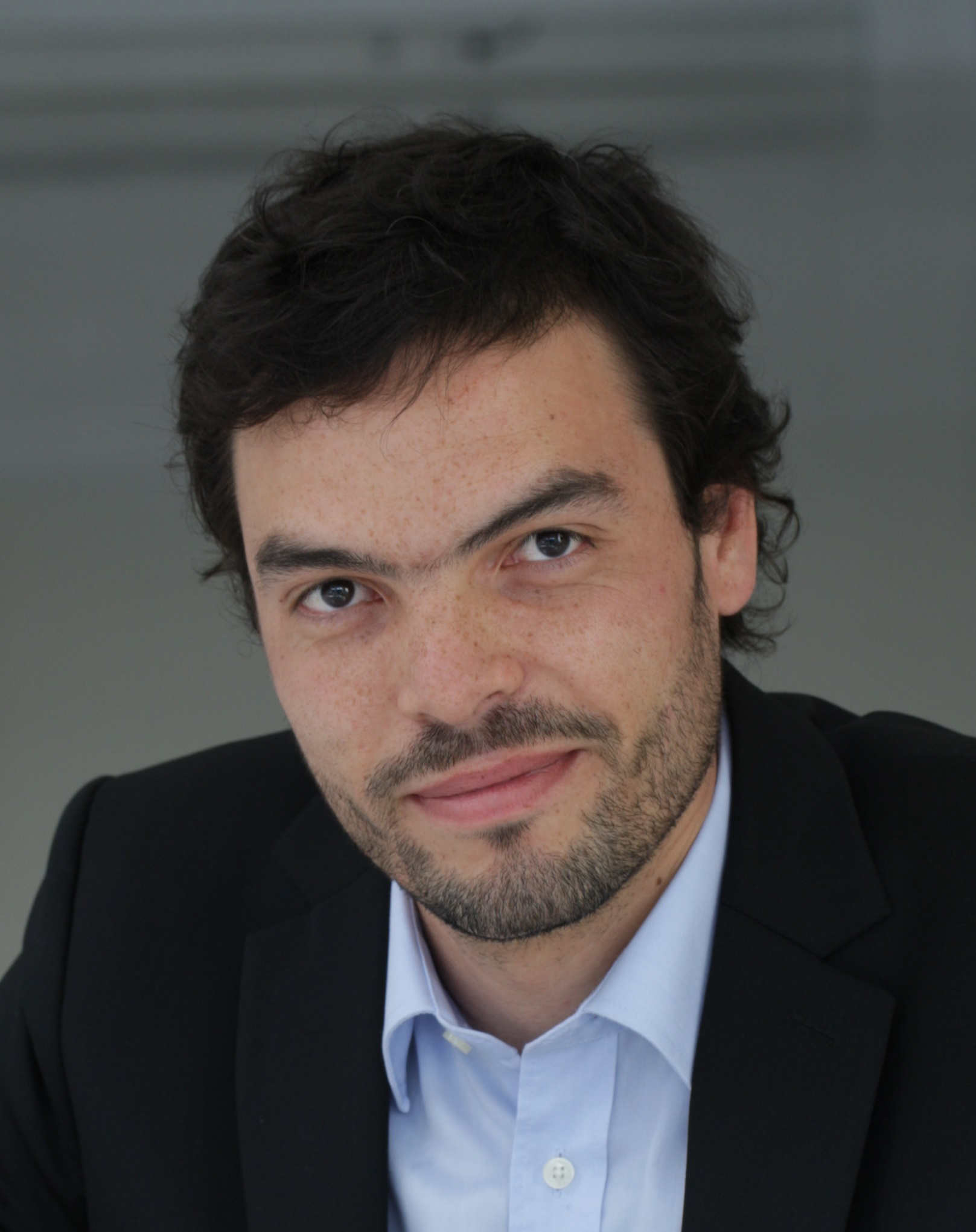}}]
{Kim Mahler} received his M.Sc. degree with honors of the EECS department from the Technical University of Berlin in 2010 and an M.A. degree from the Berlin University of Arts / University of St. Gallen in 2014. Kim is with the Wireless Communications and Networks department at the Fraunhofer Heinrich Hertz Institute and working as a researcher in projects related to vehicular communications. His research interests involve extraction of time-variant wideband multipath components and parametrization of geometry-based stochastic channel models.
\end{IEEEbiography}
\vspace*{-3\baselineskip}
\begin{IEEEbiography}[{\includegraphics[width=1in,height=1.25in,clip,keepaspectratio]{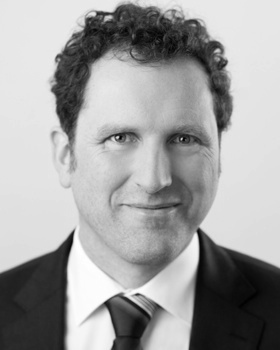}}]
{Wilhelm Keusgen} received the Dipl.-Ing. (M.S.E.E.) and Dr.-Ing. (Ph.D.E.E.) degrees from the RWTH Aachen University, Aachen, Germany, in 1999 and 2005, respectively. From 1999 to 2004, he was with the Institute of High Frequency Technology, RWTH Aachen University. Since 2004 he is heading a research group for mm-waves and advanced transceiver technologies at the Fraunhofer Heinrich Hertz Institute, located in Berlin, Germany. His main research areas are millimeter wave communications for 5G, measurement and modeling of wireless propagation channels, multiple antenna systems, and compensation of transceiver impairments. Since 2007 he also has a lectureship at the Technical University Berlin.
\end{IEEEbiography}
\vspace*{-3\baselineskip}
\begin{IEEEbiography}[{\includegraphics[width=1in,height=1.25in,clip,keepaspectratio]{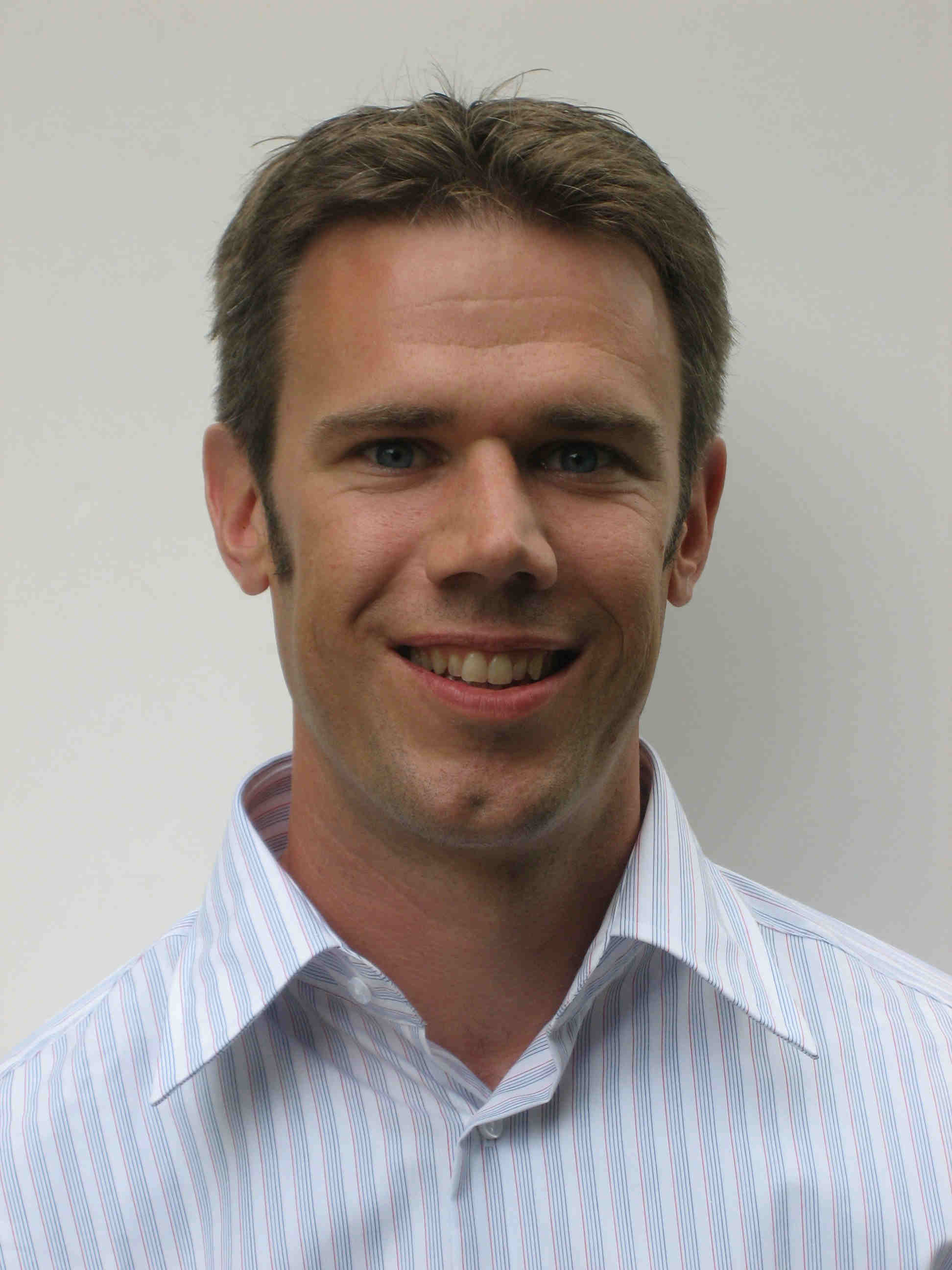}}]
{Fredrik Tufvesson} received his Ph.D. in 2000 from Lund University in Sweden. After two years at a startup company, he joined the department of Electrical and Information Technology at Lund University, where he is now professor of radio systems. His main research interests are channel modelling, measurements and characterization for wireless communication, with applications in various areas such as massive MIMO, UWB, mm wave communication, distributed antenna systems, radio based positioning and vehicular communication. Fredrik has authored around 60 journal papers and 120 conference papers, recently he got the Neal Shepherd Memorial Award for the best propagation paper in IEEE Transactions on Vehicular Technology.
\end{IEEEbiography}
\vspace*{-3\baselineskip}
\begin{IEEEbiography}[{\includegraphics[width=1in,height=1.25in,clip,keepaspectratio]{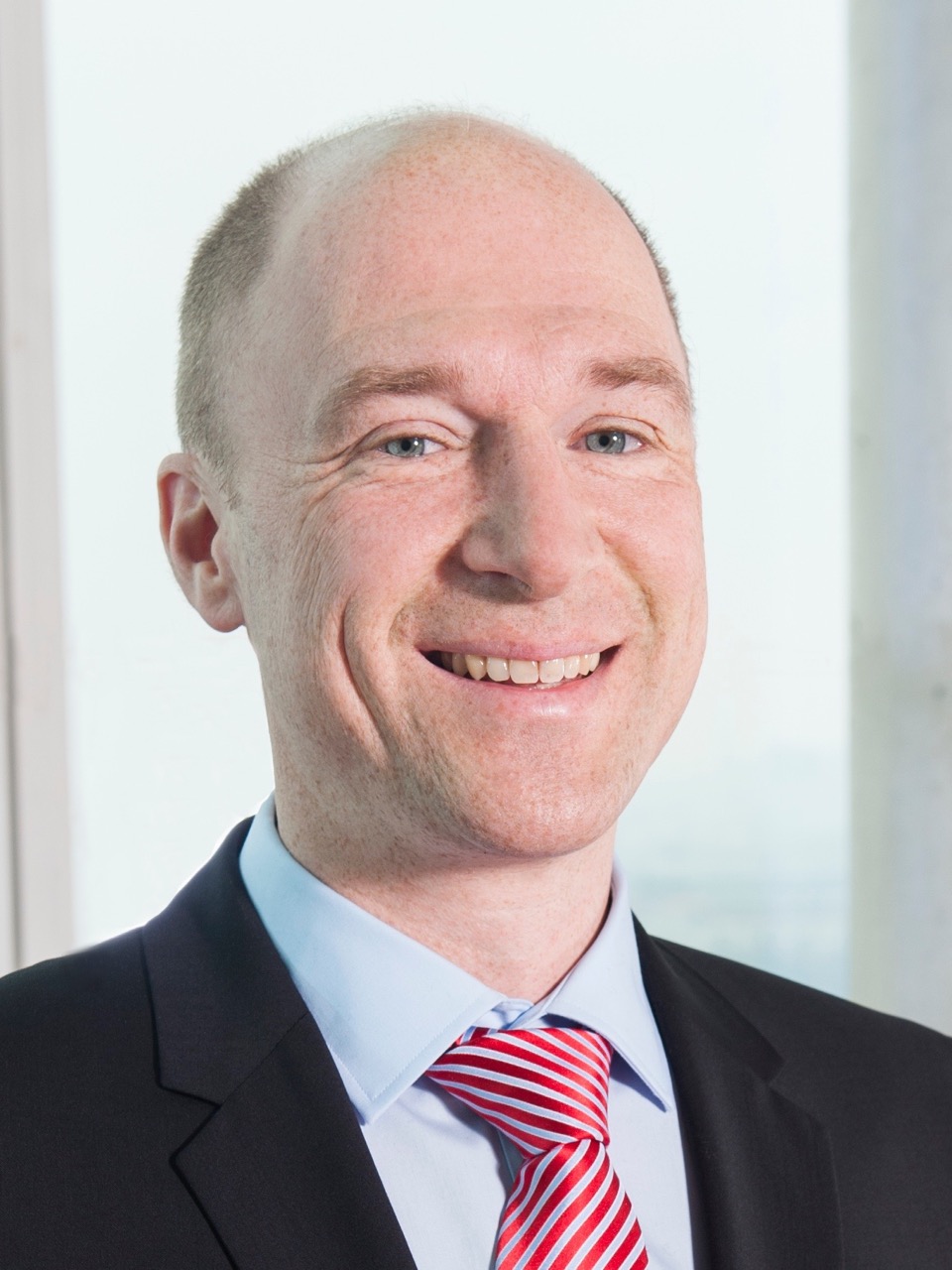}}]
{Thomas Zemen} (S'03--M'05--SM'10) received the Dipl.-Ing. degree (with distinction) in electrical engineering in 1998, the doctoral degree (with distinction) in 2004 and the Venia Docendi (Habilitation) for "Mobile Communications" in 2013, all from Vienna University of Technology.
From 1998 to 2003 he worked as Hardware Engineer and Project Manager for the Radio Communication Devices Department, Siemens Austria. From 2003 to 2015 Thomas Zemen was with FTW Telecommunications Research Center Vienna and Head of the "Signal and Information Processing" department since 2008. Since 2014 Thomas Zemen has been Senior Scientist at AIT Austrian Institute of Technology leading the research group for ultra-reliable wireless machine-to-machine communications. He is the author or coauthor of four books chapters, 32 journal papers and more than 80 conference communications. His research interests focus on reliable, low-latency wireless communications for highly autonomous vehicles, sensor and actuator networks, vehicular channel measurements and modeling, time-variant channel estimation, cooperative communication systems and interference management.
Dr. Zemen is an External Lecturer with the Vienna University of Technology and serves as Editor for the IEEE Transactions on Wireless Communications. 
\end{IEEEbiography}
\vspace*{-3\baselineskip}
\begin{IEEEbiography}[{\includegraphics[width=1in,height=1.25in,clip,keepaspectratio]{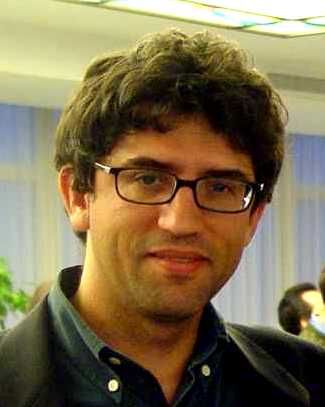}}]
{Giuseppe Caire} (S'92--M'94--SM'03--F'05) was born in Torino, Italy, in 1965. He received the B.Sc. in Electrical Engineering  from Politecnico di Torino (Italy), in 1990, the M.Sc. in Electrical Engineering from Princeton University in 1992 and the Ph.D. from Politecnico di Torino in 1994.
He has been a post-doctoral research fellow with the European Space Agency (ESTEC, Noordwijk, The Netherlands) in 1994-1995, Assistant Professor in Telecommunications at the Politecnico di Torino, Associate Professor at the University of Parma, Italy, Professor with the Department of Mobile Communications at the Eurecom Institute,  Sophia-Antipolis, France, and he is currently a professor of Electrical Engineering with the Viterbi School of Engineering, University of Southern California, Los Angeles and an Alexander von Humboldt Professor with the Electrical Engineering and Computer Science Department of the Technical University of Berlin, Germany.
He served as Associate Editor for the IEEE Transactions on Communications in 1998-2001 and as Associate Editor for the IEEE Transactions on Information Theory in 2001-2003.  He received the Jack Neubauer Best System Paper Award from the IEEE Vehicular Technology Society in 2003,  the IEEE Communications Society \& Information Theory Society Joint Paper Award in 2004 and in 2011, the Okawa Research Award in 2006, the Alexander von Humboldt Professorship in 2014, and the Vodafone Innovation Prize in 2015.
Giuseppe Caire is a Fellow of IEEE since 2005.  He has served in the Board of Governors of the IEEE Information Theory Society from 2004 to 2007, and as officer from 2008 to 2013. He was President of the IEEE Information Theory Society in 2011.
His main research interests are in the field of communications theory, information theory, channel and source coding with particular focus on wireless communications.
\end{IEEEbiography}

\end{document}